\documentclass[final,1p,times]{elsarticle}

\usepackage{url}
\usepackage{fancyvrb}

\usepackage{color}
 \usepackage{graphicx}
\usepackage{amssymb}
\usepackage{amsmath}
\newcounter{bla}

\journal{Computer Physics Communications}

\newcommand{\etal}{{\it et al. }}
\newcommand{\gray}{$\gamma$-ray}
\newcommand{\grays}{$\gamma$ rays}
\newcommand{\g}{$\gamma$}
\newcommand{\nubar}{$\overline{\nu}$}
\newcommand{\TKE}{\mathrm{TKE}}
\newcommand{\TXE}{\mathrm{TXE}}
\newcommand{\avgTKE}{\langle\mathrm{TKE}\rangle}

\newcommand{\CGMF}{$\mathtt{CGMF}$}

\newcommand{\MCNPR}{$\mathtt{MCNP6}$\textsuperscript{\textregistered}}
\newcommand{\MCNP}{$\mathtt{MCNP6}$}

\newcommand{\COH}{$\mathtt{CoH}_3$}
\newcommand{\RIPL}{$\mathtt{RIPL3}$}
\newcommand{\CGMFtk}{$\mathtt{CGMFtk}$}

\newcommand{\version}{$\mathtt 1.0$}

\definecolor{pjnotes}{RGB}{0,200,0}

\begin{document}

\begin{frontmatter}

\title{Fission Fragment Decay Simulations with the \CGMF\ Code}

\author[a]{P. Talou\corref{author}}
\author[a]{I. Stetcu}
\author[a,b]{P. Jaffke}
\author[a]{M.E. Rising}
\author[a]{A.E. Lovell}
\author[a]{T. Kawano}

\cortext[author] {Corresponding author.\\\textit{E-mail address:} talou@lanl.gov}
\address[a]{Los Alamos National Laboratory, Los Alamos, NM 87545, USA}
\address[b]{System Evaluation Division, Institute for Defense Analyses, Alexandria, VA 22311, USA}

\centerline{[ LA-UR-20-21264 ]}

%======================================================================
\begin{abstract}
The \CGMF\ code implements the Hauser-Feshbach statistical nuclear reaction model to follow the de-excitation of fission fragments by successive emissions of prompt neutrons and \grays. The Monte Carlo technique is used to facilitate the analysis of complex distributions and correlations among the prompt fission observables. Starting from initial configurations for the fission fragments in mass, charge, kinetic energy, excitation energy, spin, and parity, Y(A,Z,KE,U,J,$\pi$), \CGMF\ samples neutron and \gray\ probability distributions at each stage of the decay process, conserving energy, spin and parity. Nuclear structure and reaction input data from the \RIPL\ library are used to describe fission fragment properties and decay probabilities. Characteristics of prompt fission neutrons, prompt fission \grays, and independent fission yields can be studied consistently. Correlations in energy, angle and multiplicity among the emitted neutrons and \grays\ can be easily analyzed as a function of the emitting fragments.
\end{abstract}
%======================================================================

\begin{keyword}
nuclear fission; prompt fission neutrons and \grays; fission fragment yields

\end{keyword}

\end{frontmatter}

% \linenumbers

% Computer program descriptions should contain the following
% PROGRAM SUMMARY.

{\bf PROGRAM SUMMARY}

\begin{small}
\noindent
%{\em Manuscript Title:}                                       \\
%{\em Authors:}                                                \\
{\em Program Title:}  \CGMF~1.0                       \\
{\em Journal Reference:}                                      \\
  %Leave blank, supplied by Elsevier.
{\em Catalogue identifier:}                                   \\
  %Leave blank, supplied by Elsevier.
{\em Licensing provisions:} Open Source, BSD-3 License                  \\
  %enter "none" if CPC non-profit use license is sufficient.
{\em Programming language:} C++ \\
{\em Computer:}  any computer with a C++ compiler \\
  %Computer(s) for which program has been designed.
{\em Operating system:} any OS    \\
  %Operating system(s) for which program has been designed.
{\em RAM:} {\it TBD} bytes                                              \\
  %RAM in bytes required to execute program with typical data.
%{\em Number of processors used:}                              \\
  %If more than one processor.
%{\em Supplementary material:}                                 \\
  % Fill in if necessary, otherwise leave out.
%{\em Keywords:} Keyword one, Keyword two, Keyword three, etc.  \\
  % Please give some freely chosen keywords that we can use in a
  % cumulative keyword index.
{\em Classification:} 17.8, 17.23 \\
  %Classify using CPC Program Library Subject Index, see (
  % http://cpc.cs.qub.ac.uk/subjectIndex/SUBJECT_index.html)
  %e.g. 4.4 Feynman diagrams, 5 Computer Algebra.
{\em External routines/libraries:} none        \\
  % Fill in if necessary, otherwise leave out.
%{\em Subprograms used:}                                       \\
  %Fill in if necessary, otherwise leave out.
%{\em Catalogue identifier of previous version:}*              \\
  %Only required for a New Version summary, otherwise leave out.
%{\em Journal reference of previous version:}*                  \\
  %Only required for a New Version summary, otherwise leave out.
%{\em Does the new version supersede the previous version?:}*   \\
  %Only required for a New Version summary, otherwise leave out.
{\em Nature of problem:} Modeling of fission fragment decay \\
  %Describe the nature of the problem here.
{\em Solution method:} Monte Carlo implementation of the Hauser-Feshbach statistical theory of nuclear reactions to describe the de-excitation of fission fragments on an event-by-event basis. \\
  %Describe the method solution here.
  % \\
%{\em Reasons for the new version:}*\\
  %Only required for a New Version summary, otherwise leave out.
 %  \\
%{\em Summary of revisions:}*\\
  %Only required for a New Version summary, otherwise leave out.
{\em Restrictions:} spontaneous fission reactions for $^{238,240,242,244}$Pu and $^{252,254}$Cf; neutron-induced fission reactions from thermal up to 20 MeV for n+$^{233,234,235,238}$U, n+$^{237}$Np, and n+$^{239,241}$Pu. Binary fission only (no ternary fission). \\
  %Describe any restrictions on the complexity of the problem here.
  % \\
%{\em Unusual features:}\\
  %Describe any unusual features of the program/problem here.
  %\\
%{\em Additional comments:}\\
  %Provide any additional comments here.
   %\\
{\em Running time:} {\it TBD, but use at least 2 different scenarios as running time can vary a lot.}\\
  %Give an indication of the typical running time here.
   \\

%\begin{thebibliography}{0}
%\bibitem{1}Reference 1         % This list should only contain those items referenced in the                 
%\bibitem{2}Reference 2         % Program Summary section.   
%\bibitem{3}Reference 3         % Type references in text as [1], [2], etc.
                               % This list is different from the bibliography at the end of 
                               % the Long Write-Up.
%\end{thebibliography}
%* Items marked with an asterisk are only required for new versions
%of programs previously published in the CPC Program Library.\\
\end{small}

\newpage

% !TEX root = ../CGMF-CPC.tex
%== INTRODUCTION ==========================================================
\section{Introduction}
\label{sec:introduction}

The fission of a heavy nucleus into two or more fragments is typically accompanied by the emission of prompt neutrons and \grays. The fission fragments are produced in a certain state of deformation and intrinsic excitation energy, eventually resulting in the production of excited fission fragments that promptly evaporate neutrons and photons to reach a more stable configuration, either a ground-state or a long-lived isomeric state. The post-neutron emission fission fragments can possibly further $\beta$-decay, leading to another burst of $\beta$-delayed neutron and photon emissions.

The study of the prompt fission neutrons and photons is important to better model the nuclear fission process, constrain the collective and intrinsic configurations of the nascent fragments near the scission point, and understand the sharing of the available excitation energy between them. This study is also crucial for improving the accuracy and reliability of simulations for a wide range of applications from nuclear energy safety and efficiency to non-proliferation and stockpile stewardship missions.

Until recently, most of the nuclear data evaluation work related to prompt fission neutrons and \grays\ was limited to their average number (multiplicity) and their average energy distribution (spectrum). Even for those somewhat simple quantities, only scarce experimental data exist, limited to a few important isotopes and incident neutron energies. Phenomenological models have been developed over the years, e.g., the so-called Los Alamos model~\cite{Madland:1982}, mostly for calculating the average prompt fission neutron spectrum (PFNS). The modeling of the prompt fission \grays\ was even more limited until recently. In turn, the evaluated data on prompt fission neutron and photons present in the evaluated nuclear data libraries were very limited and/or of dubious accuracy. 

The \CGMF\ code was developed to model the de-excitation of the fission fragments on an event-by-event basis, following the successive emissions of neutrons and \grays. This approach represents a radical departure from past physics models and codes, allowing for an unprecedented set of unique predictions on distributions and correlations of neutrons, photons and fission fragments. The development of this code is accompanied by a host of modern fission experiments that have a fresh look at increasingly fine details and correlations among the vast quantity of fission data. Correlations and distributions of post-scission data are extremely useful to constrain the free parameters entering in the \CGMF\ code.

This paper reviews the physics models implemented in \CGMF, its general algorithm, and the set of model parameters used throughout. The current applicability of the code in terms of isotopes, reactions and energies is given and future directions discussed. Auxiliary nuclear data files required to run the code are also described in detail. Finally, a Python package developed to analyze \CGMF\ output files is presented. 
	%-- Introduction
% !TEX root = ../CGMF-CPC.tex
%-- GENERAL ALGORITHM & C-API -----------------------------------------------------------------
\section{General Algorithm \& C-API}

\subsection{Algorithm} \label{sec:algorithm}

The current version (ver. \version) of the \CGMF\ code was re-written significantly from previous versions (\texttt{0.x.y}) to better interface with the \footnote{\MCNPR\ and Monte Carlo N-Particle\textsuperscript{\textregistered} are registered trademarks owned by Triad National Security, LLC, manager and operator of Los Alamos National Laboratory. Any third party use of such registered marks should be properly attributed to Triad National Security, LLC, including the use of the designation as appropriate. For the purposes of visual clarity, the registered trademark symbol is assumed for all references to MCNP within the remainder of this paper.}\MCNPR\ transport code. It was also developed such that the \texttt{cgmfEvent} C++ class inherits from the \texttt{cgmEvent} class, which handles non-fission events such as inelastic scattering. This organization makes it straightforward to improve physics models or computational routines for both fission and non-fission event simulations. However note that the \CGMF\ code described in this paper is intended solely for fission event simulations, and no discussion or support is provided for non-fission event simulations at this point. 

The \CGMF/\MCNP\ interface is implemented in the \texttt{cgmfEvents.cpp/h} files. The simulation of a fission event is triggered by the instruction:
\begin{eqnarray}
\texttt{cgmf\_genfissevent (int ZAIDt, double En, double time)} \nonumber 
\end{eqnarray}
\noindent
where \texttt{ZAIDt} represents the index number (1000$\times$Z+A) of the target nucleus, \texttt{En} is the energy of the incident neutron (in MeV), and \texttt{time} is a time stamp used to carry information about delayed isomeric decays. In the case of spontaneous fission, \texttt{En} is set to zero, and \texttt{ZAIDt} implicitly represents the compound nucleus \texttt{ZAIDc}.

This instruction leads to the instantiation of a \texttt{cgmfEvent}, only after the input given by the user is first validated against the range of isotopes, reactions and energies allowed in the current version of the code. The \texttt{cgmfEvent::initialization} function is then called, for the first event only. During this initialization phase, various data files needed by \CGMF\ to complete a calculation will be read: ground-state nuclear masses and deformations, discrete level data, level density parameter and temperature systematics, etc. This is done for all nuclei in the nuclear chart, so that any new instantiation of a \texttt{cgmfEvent}, for a different isotope or energy for instance, does not require new files to be read. Those auxiliary data files are discussed at more length in Section~\ref{sec:data}.

Once all input data are read, array dimensions set, and the random number generator seed initialized, a \CGMF\ calculation follows this sequence:

\begin{enumerate}
\item Define the compound nucleus mass, charge and excitation energy based on the reaction specified by the user input;
\item If pre-fission neutron emission is energetically allowed, sample the pre-fission neutron energy spectrum (pre-calculated) to select the energy or energies of the neutron(s) emitted, and determine the mass, charge and excitation energy of the residual nucleus that will undergo fission;
\item Sample the fission fragment mass distribution for this initial configuration, and select the masses for the heavy ($A_H$) and light ($A_L$) complementary fragments;
\item Sample the fission fragment charges for the light ($Z_L$) and heavy ($Z_H$) fragments;
\item Compute the $Q_f$-value of the fission reaction based on tabulated nuclear masses;
\item Sample the total kinetic energy (TKE) based on the type of fission reaction (neutron-induced or spontaneous fission), the fissioning nucleus, and the particular fragmentation $(A_H,Z_H)+(A_L,Z_L)$.
\item Infer the total excitation energy (TXE) available for neutron and \g\ emissions based on $Q_f$ and TKE;
\item Share TXE between the two fragments and compute the initial excitation energies $U_L$ and $U_H$ for the light and heavy fragments, respectively;
\item Sample the spin distributions for both fragments at their given initial excitation energies to select initial spins $J_L$ and $J_H$. Parities are selected randomly between negative and positive;
\item Start the de-excitation loop for the light fragment:
\begin{itemize}
\item Compute the energy-dependent neutron transmission coefficients $T_{lj}(\epsilon)$ using optical model calculations;
\item Compute the \gray~transmission coefficients from a GDR representation;
\item For the specific initial configuration $(U_i,J_i,\pi_i)$, select a decay channel, either neutron or \g, based on their emission probabilities;
\item Record the decay path $(i\rightarrow f)$, and point to the final state $(U_f,J_f,\pi_f)$ in the residual nucleus;
\item Repeat until a ground-state or isomeric state is reached.
\end{itemize}
\item Perform a similar calculation loop for the heavy fragment;
\item Return the result of the calculations for both fragments, prompt neutrons and \grays.
\end{enumerate}

%-- CGMF flowchart
\begin{figure}[ht]
\centerline{\includegraphics[width=0.75\columnwidth]{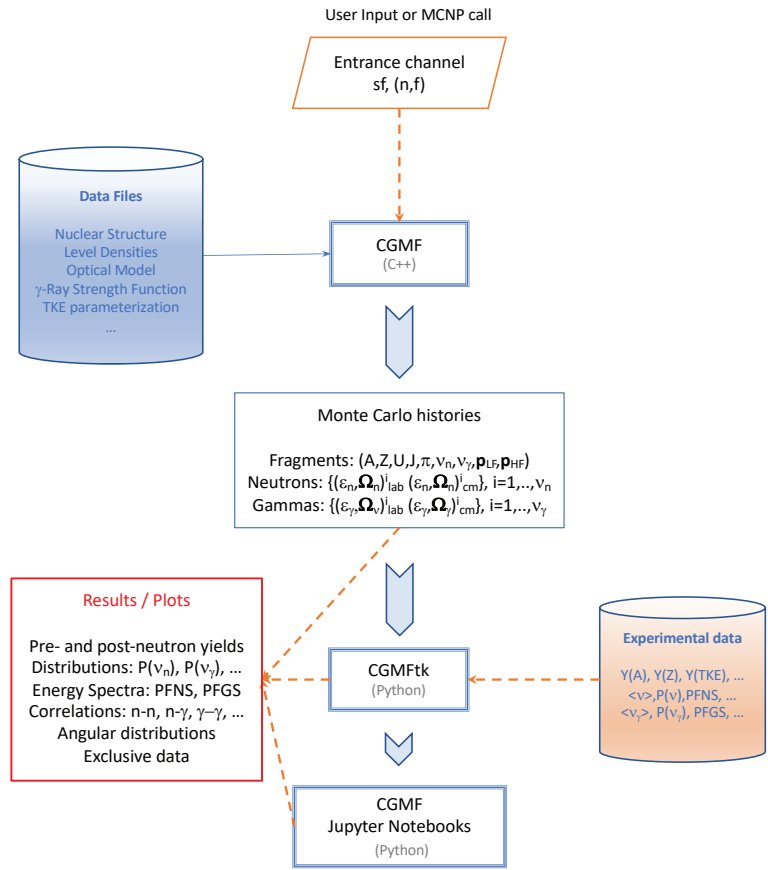}}
%\centerline{\includegraphics[width=0.82\columnwidth]{figs/flowchart.png}}
\caption{\label{fig:flowchart}Flowchart of the \CGMF\ code. \CGMF\ makes use of the \RIPL\ database for a wide range of input parameters. Monte Carlo histories are stored in an ASCII output file, which can then be conveniently analyzed using the Python routines implemented in the \CGMFtk\ toolkit. Some of those routines can also be used to compare calculated results with experimental data for all kinds of fission observable (spectrum, multiplicity, ...). Jupyter notebooks are also commonly used to facilitate this analysis stage.}
\end{figure}

The results consist in a {\it history} ASCII data file that contains all {\it N} Monte Carlo events run by the user. For each fission event, this data file contains the complete information needed for the reconstruction of the decay sequence starting with the initial fragment mass, charge, excitation energy, kinetic energy, spin and parity, followed by the characteristics in energy and momenta for all neutrons and \grays\ emitted in that event. An example of such an output file is given in Section~\ref{sec:running}. A set of Python routines has been written in the \CGMFtk\ package, discussed in Section~\ref{sec:CGMFtk}, which can be used to efficiently analyze and plot all sorts of averages, distributions, and correlations between the emitted particles and the fragments.

A flowchart describing some of the key elements of a \CGMF\ run, its input and output files, and the result analysis process is shown in Fig.~\ref{fig:flowchart}.

%-----------------------------------------------------------------------------------
\subsection{Generating fission yields} \label{sec:fissionYields}
%-----------------------------------------------------------------------------------

Sometimes it is useful to generate the initial {\it pre-neutron} emission fission fragment yields, without having to calculate the full decay phase of neutron and \gray\ evaporation. \CGMF\ provides this option, and the initial fission fragment distributions in mass, charge, kinetic energy, excitation energy, spin and parity, Y(A,Z,KE,U,J,$\pi$), can be simply generated using:
\begin{eqnarray}
\texttt{cgmf\_genfissyields (int ZAIDt, double En, int nevents)}, \nonumber
\end{eqnarray}
where \texttt{ZAIDt} is the ZAID number (1000$\times$Z+A) of the target nucleus, \texttt{En} the energy of the incident neutron (set to zero in the case of spontaneous fission), and \texttt{nevents} is the number of fission events to be generated.

%-----------------------------------------------------------------------------------
\subsection{C-API} \label{sec:CAPI}
%-----------------------------------------------------------------------------------

The Application Programming Interface (API) that comes with \CGMF\ provides a set of routines for running the code, and manipulating and retrieving its results. Table~\ref{table:routines} provides a very partial list of some commonly used routines. This API was created to facilitate the simulation of fission events in transport codes such as \MCNP. When used as a standalone code, \CGMF\ routines can be accessed directly instead. The command line instruction to run \CGMF\ is:
\bigskip

\texttt{> cgmf.x -i ZAIDt -e incidentEnergy -n numberOfMonteCarloEvents}

\bigskip
\noindent
Further details are provided in Section~\ref{sec:running}.
 
%-- TABLE: List of common CGMF routines provided in the API ---------------------------------------------------
\begin{table}[h]
\centering
\setlength{\tabcolsep}{2pt}
\caption{Common \CGMF\ routines provided in the API. If `\texttt{(...)}' is used, input parameters are not given explicitly. \label{table:routines} }
\begin{tabular}{p{5.5cm}p{7.5cm}}
\hline \hline
Routine call & Action \\
\hline
\texttt{void cgmf\_genfissevent (...)} & Generates one fission event for a specific target nucleus, incident neutron energy and time coincidence window\\
\texttt{void cgmf\_genfissyields (...)} & Returns initial (A,Z,KE,U,J,$\pi$) for a specified number of fission events for a given target nucleus and  incident neutron energy \\
\texttt{int cgmf\_getnnu()} & Returns the number of neutrons emitted in this fission event \\
\texttt{double cgmf\_getnerg (int i)} & Returns the energy of the $i^{\rm th}$ neutron emitted in the laboratory\\
\texttt{double cgmf\_getndircosu(int i)} & Returns the first directional cosine of the $i^{\rm th}$ emitted neutron\\
\texttt{double cgmf\_getntme(int i)} & Returns the time of emission of the $i^{\rm th}$ neutron\\
\texttt{int cgmf\_getlfmass()} & Returns the light fragment mass\\
\texttt{double cgmf\_getlfke()} & Returns the light fragment pre-neutron emission kinetic energy\\
\texttt{int cgmf\_getlfnnu()} & Returns the number of neutrons emitted from the light fragment\\
\texttt{int cgmf\_getlfgnu()} & Returns the number of photons emitted from the light fragment\\
\texttt{double cgmf\_getlfdircosv()} & Returns the second directional cosine of the light fragment trajectory\\
\hline\hline
\end{tabular}
\end{table}

		%-- General algorithm of the code
% !TEX root = ../CGMF-CPC.tex

%== PHYSICS MODELS =========================================================
\section{Physics Models}
\label{sec:models}

The de-excitation of the fission fragments by neutron and \gray\ emissions is modeled using the Hauser-Feshbach statistical model of nuclear reactions~\cite{Hauser:1952}. In this model, the compound nucleus formed in a particular excitation energy, spin and parity state, $(U,J,\pi)$, decays into different channels according to probabilities, or widths. In the case of fission fragments produced in low-energy fission reactions, only neutron and \gray\ emission probabilities are significant, while charged-particle emissions are strongly hindered due to the Coulomb barrier. This is not to be confused with ternary fission processes, where charged particles can be emitted in a dynamical process, as opposed to statistical evaporation. In its current version, \CGMF\ models binary fission only, but does account for the emission of pre-scission neutrons and multi-chance fission.

% !TEX root = ../CGMF-CPC.tex
%-- PHYSICS MODELS: Fission Fragment Yields ------------------------------------------------------------------------
\subsection{Fission Fragment Mass, Charge and Kinetic Energy Distributions}
\label{sec:yields}

Any given fission reaction produces a wide distribution of fission fragments in mass, charge and kinetic energy, reflecting the complexity and variety of configurations that the parent nucleus can assume before splitting into two or more lighter nuclei. The pre-scission phase of the fission process has been the object of numerous theoretical studies, including a macroscopic-microscopic approach~\cite{Randrup:2011,Moller:2015,Mumpower:2020}, which treats the nucleus as a quantum charged liquid drop that deforms due to Coulomb repulsion and nuclear attractive forces, and a purely microscopic approach~\cite{Schunck:2014,Bulgac:2016} which starts from phenomenological descriptions of the nucleon-nucleon forces. Both approaches have merits and limitations, but it is only very recently that predictions have been made on primary fission fragment yields~\cite{Regnier:2016,Sierk:2017,Usang:2017,Regnier:2019}.

By default, \CGMF~implements a three-Gaussian model to represent pre-neutron fission fragment mass distributions, as follows:
\begin{eqnarray}
Y(A;E_n) = G_0(A)+G_1(A)+G_2(A),
\label{eq:YA}
\end{eqnarray}
where $G_0$ corresponds to a symmetric mode,
\begin{eqnarray}
G_0(A) = \frac{W_0}{\sigma_0\sqrt{2\pi}}{\rm exp}\left( -\frac{(A-\overline{A})^2}{2\sigma_0^2} \right),
\end{eqnarray}
and $G_1$ and $G_2$ are the two asymmetric modes
\begin{eqnarray}
G_{1,2}(A) = \frac{W_{1,2}}{\sigma_{1,2}\sqrt{2\pi}}\left[  {\rm exp}\left( -\frac{(A-\mu_{1,2})^2}{2\sigma_{1,2}^2}\right) + {\rm exp}\left( -\frac{\left(A-(A_p-\mu_{1,2})\right)^2}{2\sigma_{1,2}^2}\right)\right].
\end{eqnarray}
Here, $\overline{A}=A_p/2$, with $A_p$ the mass of the parent fissioning nucleus, which can differ from the original compound nucleus $(Z_c,A_c)$ if pre-fission neutrons are emitted. The parameters $\mu_i$ are the means, $W_i$ are the weights, and $\sigma_i$ are the widths of the Gaussian modes. In the case of pre-fission neutron emission, the parameters are chosen for the resulting parent fissioning nucleus (i.e. $A_p = A_0 - \nu_\mathrm{pre}$). 

\begin{figure}[ht]
\centerline{
\includegraphics[width=0.48\columnwidth]{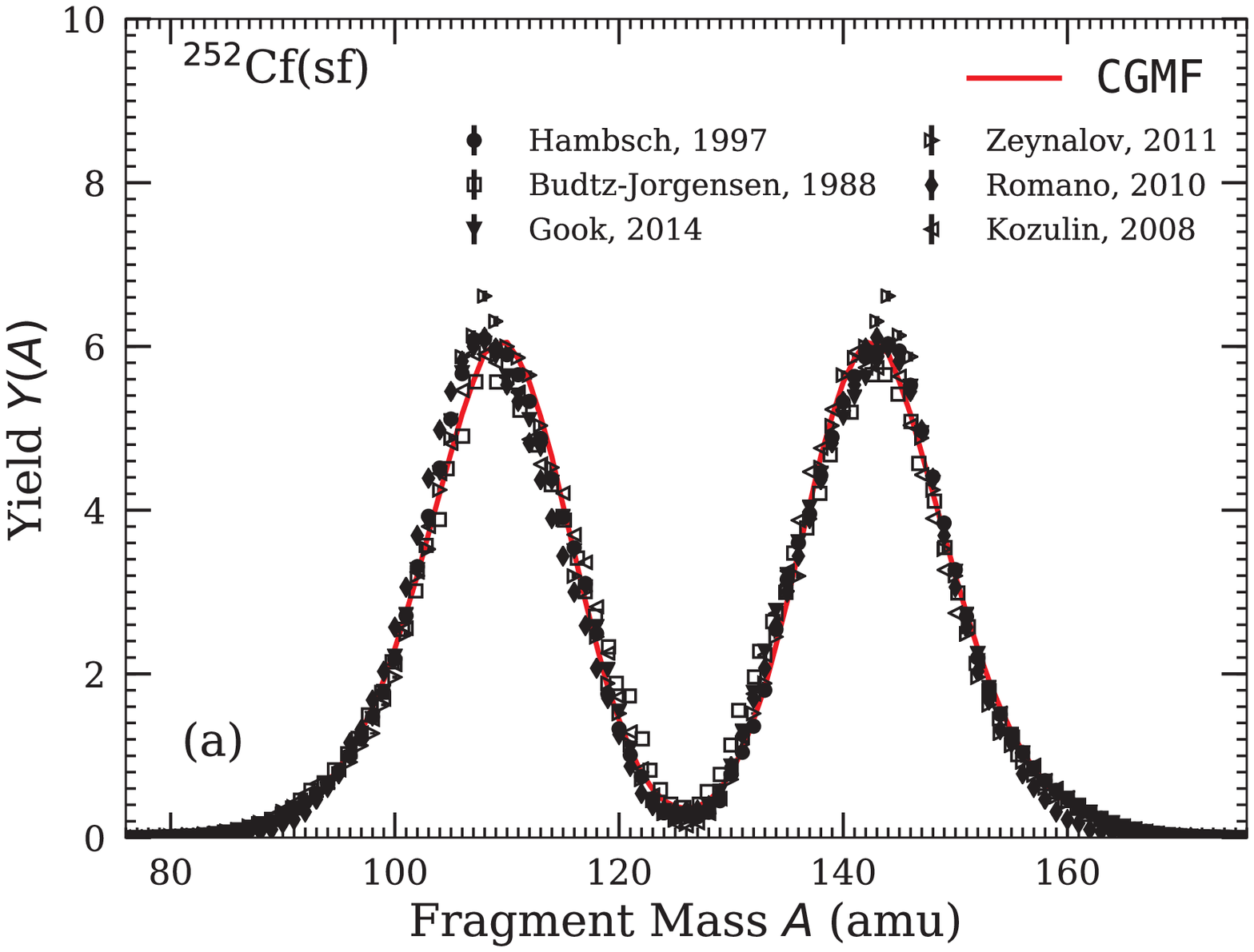}
\includegraphics[width=0.48\columnwidth]{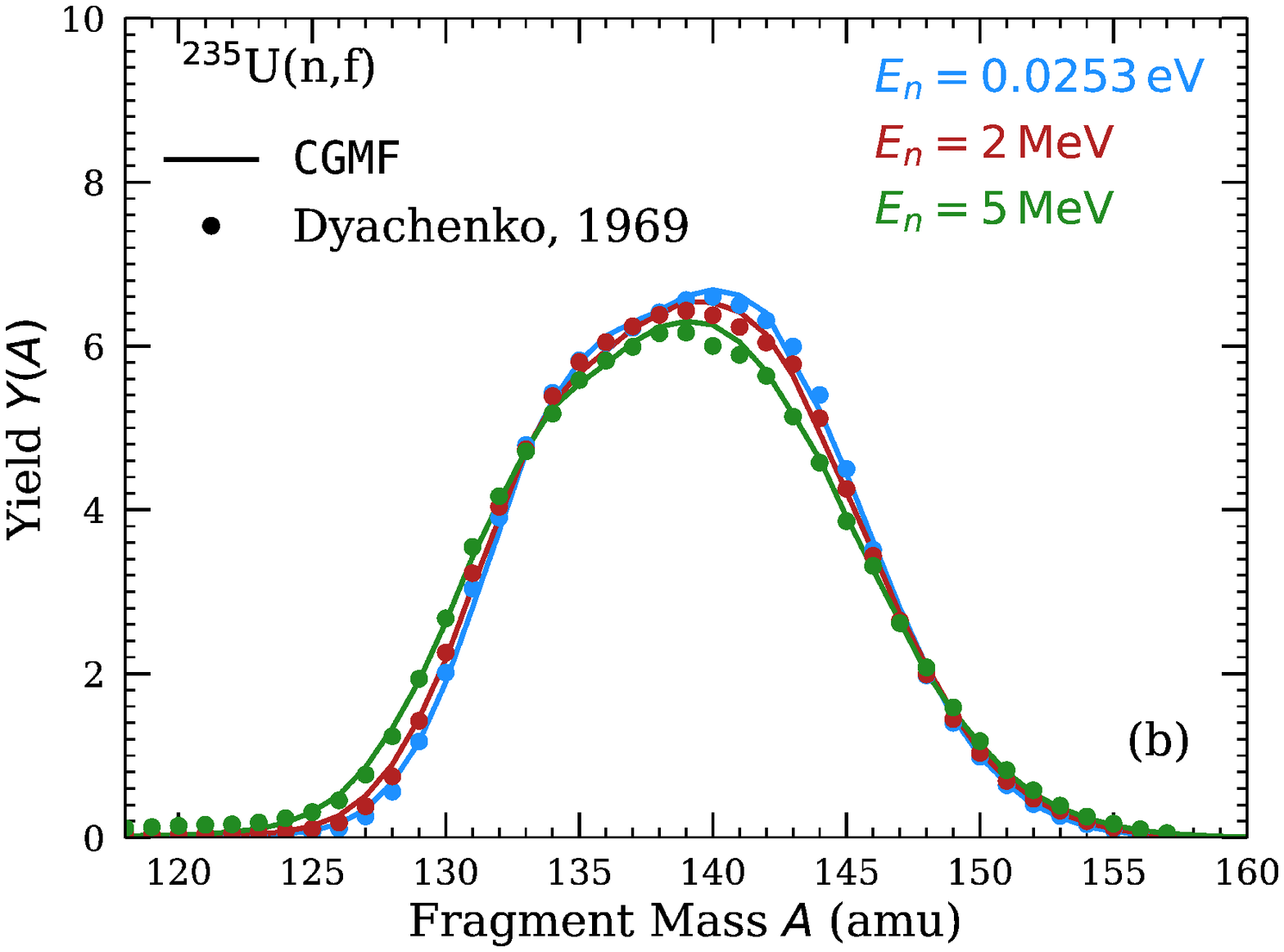}
}
\caption{\label{fig:YA}Fission fragment mass yield Y(A) for (a) the spontaneous fission of $^{252}$Cf (experimental data are from~\cite{Hambsch:1997,Budtz-Jorgensen:1988,Gook:2014,Zeynalov:2011,Romano:2010,Kozulin:2008}), and (b) the neutron-induced fission of $^{235}$U for several incident neutron energies (experimental data are from~\cite{Dyachenko:1969}). The lines represent \CGMF\ calculations for 1M events, while the markers are experimental data.}
\end{figure}

In \CGMF~, the Gaussian mode parameters must have an energy dependence. The means and widths are linear in the incident neutron energy
\begin{eqnarray}
\mu_i = \mu_i^{(0)} + \mu_i^{(1)}E_n \hspace{20mm} \sigma_i = \sigma_i^{(0)} + \sigma_i^{(1)}E_n.
\end{eqnarray}
The constants were determined from a least-squares fit of experimental data at a given incident energy, typically thermal neutrons. The energy dependence for the weights is given by a Fermi function
\begin{eqnarray}
W_i = \frac{1}{1+\exp[(E_n-W_i^{(0)})/W_i^{(1)}]}.
\end{eqnarray}
The weight of the symmetric Gaussian $W_0$ is determined from the conservation equation $W_0 = 2 - 2W_1 - 2W_2$. Similarly, a matching energy was used to reduce the number of fitted parameters by setting the ratio of $W_i^{(0)}/W_i^{(1)}$. The fission fragment mass yields obtained for $^{252}$Cf(sf) and for $^{235}$U(n,f) are shown in Fig.~\ref{fig:YA}.

%-- FIGURE: Y(A) for Cf252(sf)

\begin{figure}[ht]
\centerline{\includegraphics[width=0.75\columnwidth]{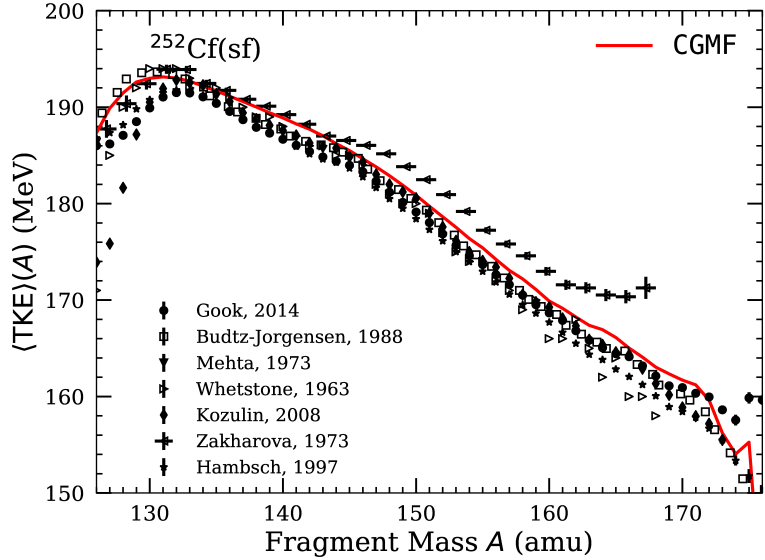}}
\caption{\label{fig:TKEA} (Color online) Average pre-neutron emission fission fragment total kinetic energy $\avgTKE$ as a function of mass $A$. The markers indicate experimental data~\cite{Gook:2014,Budtz-Jorgensen:1988,Mehta:1973,Whetstone:1963,Kozulin:2008,Zakharova:1973,Hambsch:1997}, while the solid red line indicates the parameterization used in \CGMF. The \CGMF\ simulations used 1M events.}
\end{figure}

The mass-dependent $\TKE$ distributions were taken to be Gaussians centered around a mean value $\mu_\TKE(A_H)$, with a width $\sigma_\TKE(A_H)$
\begin{eqnarray}
\label{eq:YTKE}
Y(\TKE;A_H;E_n) = \frac{1}{\sigma_\TKE(A_H)\sqrt{2\pi}}\exp\bigg[\frac{(\TKE-\eta\times\mu_\TKE(A_H))^2}{2\sigma_\TKE^2(A_H)}\bigg].
\end{eqnarray}
The mean and width depend on the heavy fragment mass $A_H$, but do not yet depend on the incident neutron energy. Both the mean and width are parameterized as polynomial fits
\begin{eqnarray}
\mu_\TKE(A_H) = \displaystyle\sum_{i=0}^N a_i (A_H - A_m)^i \hspace{20mm} \sigma_\TKE(A_H) = \displaystyle\sum_{i=0}^N b_i (A_H - A_m)^i.
\end{eqnarray}
The value of $N$ is typically between $4$ and $6$ and $A_m$ is the mass resulting in the maximum $\mu_\TKE$. The coefficients $a_i$ and $b_i$ are determined with a least-squares fit to experimental data. To avoid sampling unphysical $\TKE$ values for masses exceeding the range of experimental data, a maximum heavy fragment mass $A_m$ value is also provided, above which \CGMF\ defaults to $\TKE=140\,\mathrm{MeV}$ and $\sigma_\TKE$=7 MeV. An example of $\langle$TKE$\rangle$(A) is shown in Fig.~\ref{fig:TKEA} for $^{252}$Cf(sf).

%-- FIGURE: <TKE>(A) for Cf252(sf)

We scale $\mu_\TKE(A_H)$ to correspond to the correct average $\TKE$ by the factor
\begin{eqnarray}
\eta = \frac{\avgTKE(E_n)}{\displaystyle\sum_{A\geq\bar{A}} \bigg[ \mu_\TKE(A) \cdot Y(A;E_n)\bigg]}.
\end{eqnarray}
The average $\TKE$, $\avgTKE(E_n)$ depends on the incident neutron energy and is parameterized by two linear functions
\begin{equation}
\label{eq:TKEE}
  \avgTKE(E_n) = \begin{cases}
    \kappa^{(0)} + \kappa^{(1)}E_n, & \text{if $E_n \leq E_c$}\\
    \rho^{(0)} + \rho^{(1)}E_n, & \text{if $E_n > E_c$}
  \end{cases}
\end{equation}
where $E_c$ is a critical energy that can define a slope change in $\avgTKE(E_n)$, typically $E_c \approx 1\,\mathrm{MeV}$, as observed in some recent experimental data~\cite{Duke:2015}. For fission reactions not displaying this slope change, $E_c = 0\,\mathrm{MeV}$. Values for $\kappa^{(0)}$, $\kappa^{(1)}$, $\rho^{(0)}$, $\rho^{(1)}$, and $E_c$ are determined from least-squares fit to experimental data. One of the parameters is determined from continuity at $E_n = E_c$ and we allow for an overall scaling of $\avgTKE(E_n)$ to match experimental prompt neutron multiplicity \nubar~data, as uncertainties on TKE (not better than about 1~MeV) typically dominate those of \nubar~\cite{Jaffke:2018,Randrup:2019}.

\begin{figure}[ht]
\centerline{\includegraphics[width=0.75\columnwidth]{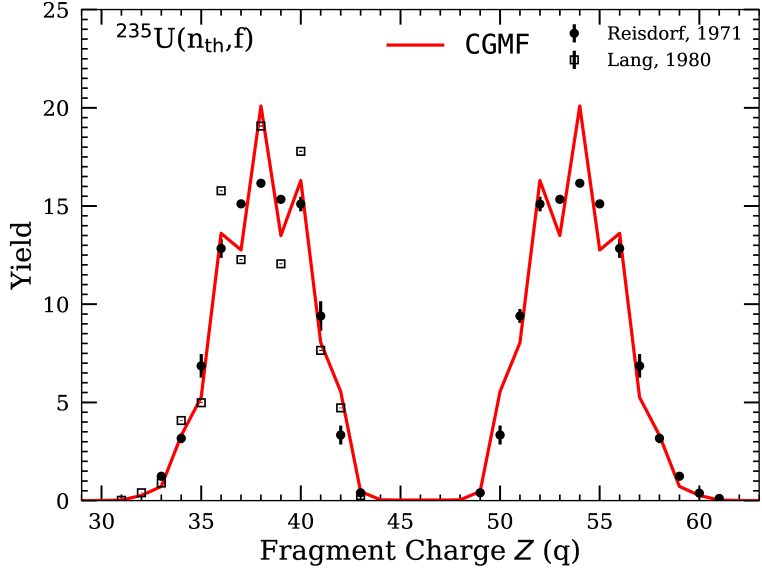}}
\caption{\label{fig:YZ}Fission fragment charge yield Y(Z) for the thermal-neutron-induced fission of $^{235}$U. Experimental data is from Ref.~\cite{Reisdorf:1971,Lang:1980}.}
\end{figure}

The charge distributions $Y(Z;A,E_n)$ follow a Gaussian form, as described by the Wahl systematics~\cite{Wahl:2002}. The means are given by the unchanged charge distribution and the so-called charge polarization, which depends on the fissioning nucleus and the incident neutron energy. The widths are a function of the fragment mass, which comes from energy-dependent fits to experimental data. Finally, shell effects are included by increasing (decreasing) the probability for even-even (odd-odd) fragments. The shell effects are also a function of fragment mass, fissioning nucleus, and incident neutron energy. The full description of $Y(Z;A,E_n)$ can be found in Ref.~\cite{Wahl:2002}. An example of the resulting total charge yield distribution Y(Z) is shown in Fig.~\ref{fig:YZ} for the thermal-neutron-induced fission of $^{235}$U where the odd-even effects are particularly visible.

%-- FIGURE: Y(Z) for U235(n,f) at thermal

If neutrons are emitted prior to fission, the fissioning nucleus is formed with a residual excitation energy smaller than the initial excitation energy. In this case, an ``equivalent" incident neutron energy is defined as the neutron energy that would produce the $(A_0-\nu_{pre})$ fissioning nucleus, with $\nu_{pre}$ pre-fission neutrons, at the same residual excitation energy. Hence, $E_n$ becomes
\begin{eqnarray}
E_n^\prime = E^* - \displaystyle\sum_{\nu=1}^{N_\mathrm{pre}}\bigg[ \epsilon^{(\nu)}_\mathrm{pre} + S_n(Z_c,A_c+1-\nu)\bigg],
\end{eqnarray}
where $E^* = E_n + S_n(Z_c,A_c)$ represents the excitation energy of the initial compound nucleus, $\epsilon^{(\nu)}_\mathrm{pre}$ represents the $\nu^\mathrm{th}$ pre-fission neutron energy (in CM frame), and $S_n(Z,A)$ is the one-neutron separation energy of a particular nucleus.
The same equivalent incident energy is used in the mass, $\TKE$, and charge distributions. In the current version of the code, we impose that $E^*$ be greater or equal than the fission barrier height in the $(A_c-\nu_{pre})$ nucleus, and therefore neglect any (small) sub-barrier fission events.

The complete fission fragment yields are reconstructed as
\begin{eqnarray}
Y(A,Z,\TKE;E_n) = Y(A;E_n) \times Y(\TKE;A_H,E_n) \times Y(Z;A,E_n)
\end{eqnarray}
First, \CGMF~ will sample from the above distribution for a pair of fragments, ($A_L,Z_L$) and ($A_H,Z_H$), along with a total kinetic energy $\TKE$. Next, we determine the total excitation energy $\TXE$ and share it between the two fragments as described next.

% !TEX root = ../CGMF-CPC.tex
%-- PHYSICS MODELS: Excitation Energy Distributions -----------------------------------------------------------------
\subsection{Excitation Energy Distributions}
\label{sec:energy}

The total excitation energy is determined using energy conservation of the sampled fission fragments. Assuming a pair of fission fragments $(A_L,Z_L)$ and $(A_H,Z_H)$, along with a sampled total kinetic energy $\TKE(A_H)$, the total excitation energy $\TXE$ is given by
\begin{eqnarray}
\TXE = [E^* + M(A_p,Z_p)c^2 - M(A_L,Z_L)c^2 - M(A_H,Z_H)c^2] - \TKE(A_H),
\label{eq:Econserve}
\end{eqnarray}
where the term in brackets represents the Q-value of the fission reaction. The Q-value is composed of the excitation energy $E^*$ of the parent fissioning nucleus ($A_p$,$Z_p$) and the mass difference between it and the sampled fragments. The masses $M(A,Z)$ are taken from the Audi-Wapstra 2012 table~\cite{Audi:2012} for nuclei with experimentally known masses, or calculated with the Finite-Range Droplet Model (FRDM) by M\"oller {\it et al.}~\cite{Moller:1995}. The excitation energy of the fissioning nucleus is given by $E^*=E_{\rm inc}+S_n$, in the case of neutron-induced fission, where $E_{\rm inc}$ is the incident neutron energy and $S_n$ the neutron separation energy of the target nucleus. In the case of spontaneous fission, $E^*$ is zero. Once TXE is known, it still needs to be distributed among the two fragments.

Several approaches exist in the literature~\cite{Litaize:2015,Schmidt:2010,Becker:2013,Morariu:2012} for sharing the excitation energy between the two fission fragments. All rely on some assumptions about the fission fragment configurations near the scission point. To first order, the amount of excitation energy given to a fragment is reflected in the number of prompt neutrons emitted by that fragment. Thus, as the average $\TXE$ is a good indicator of the average total prompt neutron multiplicity \nubar, the sharing of this energy is a good indicator of the prompt neutron multiplicity for a particular fragment \nubar~$(A)$. In \CGMF, we introduce a ratio of nuclear temperatures $R_T$, given by
\begin{eqnarray}
R_T^2 = \frac{T_L^2}{T_H^2} \approx \frac{U_L a_H(U_H)}{U_H a_L(U_L)}.
\label{eq:RT}
\end{eqnarray}
In Eq.~\ref{eq:RT}, the approximation assumes a constant temperature and Fermi gas model for the level density in order to relate the nuclear temperature $T$ to the excitation energy $U$ and level density parameter $a$. The level density parameters are energy-dependent $a \equiv a(U)$, so Eq.~\ref{eq:RT} is solved iteratively. In \CGMF, $R_T$ can be mass-dependent, $R_T \equiv R_T(A_H)$, and this function is fitted to match the available \nubar~$(A)$ data.

Using a ratio of nuclear temperatures, as in \CGMF, is theoretically motivated by the deformation of the fission fragments. The heavy fragments, typically formed near the $Z=50$ and $N=82$ closed shells, are expected to be less deformed than their light partners. Thus, the light fragments are expected to acquire more of the total excitation energy, meaning $R_T > 1$. The $R_T(A_H)$ in \CGMF\ follows this trend and provides good agreement with the sawtooth shape of \nubar~$(A)$ data, while having little impact on most other fission observables. Currently, this parameterized $R_T(A_H)$ function does not depend on the incident neutron energy, meaning that the sawtooth merely scales with the total prompt neutron multiplicity in contrast to some experimental data~\cite{Mueller:1984,Naqvi:1986}. Additional work is ongoing to provide a more robust model for the excitation energy for a future version. 

%\PJ{This model~\cite{Jaffke:2019} attempts to combine a distribution in scission shapes, which is based on calculated deformation energies, and shares the remaining intrinsic excitation energy via maximum entropy~\cite{Morariu:2012}.} \textcolor{red}{We can include this sentence, but it could take a long time to get this paper finished, so I'd rather remove it. Thoughts? }

% !TEX root = ../CGMF-CPC.tex
%-- PHYSICS MODELS: Initial Spin and Parity Distributions -------------------------------------------------------------
\subsection{Initial Spin and Parity Distributions}
\label{sec:spin}

The spin of the fragments also follows a conservation rule

\begin{equation}
\vec J_1 +\vec J_2+\vec l=\vec J
\end{equation}
where $\vec J_1$ and $\vec J_2$ are the fission fragment spins, $\vec J$ is the total angular momentum of the fissioning nucleus, and $\vec l$ is the relative orbital angular momentum between the two fragments. In the current version of CGMF, $\vec J_1$ and $\vec J_2$ follow a Gaussian distribution around a mean value that is chosen to best reproduce some of the observed prompt photon characteristics. The relative orbital angular momentum $\vec l$ is left free, so there is no correlation between $\vec J_1$ and $\vec J_2$ at this point. This question will be revisited in future versions of the code. Also, negative and positive parities are chosen to be equally probable, so the spin and parity distribution in the fragments is

\begin{equation}
\rho(J,\pi)=\frac{1}{2}(2J+1)\exp\left[-\frac{J(J+1)}{2B^2(Z,A,T)}\right],
\end{equation}
where $B$ is defined in terms of the fragment temperature $T$ and the ground-state moment of inertia $\mathcal{I}_0(A,Z)$:
\begin{equation}
B^2(Z,A,T)=\alpha\frac{\mathcal{I}_0(A,Z)T}{\hbar^2}.
\label{eq:b2}
\end{equation}
In Eq. (\ref{eq:b2}), $\alpha$ is an adjustable parameter that is used to globally adjust the competition between neutrons and photons, so that, for neutron-induced reactions, $\bar\nu$ is reproduced as a function of neutron  energy. At the same time, $\alpha$ is chosen so that for thermal neutron-induced and spontaneous fission reactions, prompt $\gamma$ fission data is also reproduced. The energy dependence of $\alpha(E_n)$ is linear and has been fitted using limited experimental total prompt \gray\ energy measurements over a range of incident neutron energies~\cite{Frehaut:1982}, as well as \gray\ multiplicity data~\cite{Chyzh:2014,Oberstedt:2015}.

%-- Initial average fragment spin as a function of the fragment mass
\begin{figure}[ht]
\centerline{\includegraphics[width=0.65\columnwidth]{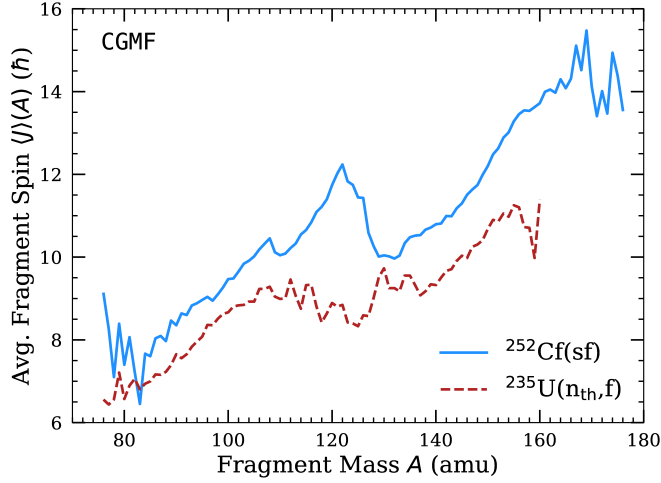}}
\caption{\label{fig:avJA} (Color online) Average pre-neutron emission fission fragment spin $\langle J\rangle$ as a function of mass $A$ for two reactions supported in \CGMF\ : $^{252}$Cf(sf) and $^{235}$U(n$_\mathrm{th}$,f). \CGMF\ Simulations used 1M events.}
\end{figure}

Typical values calculated as a function of the light and heavy fragment masses are shown in Fig. \ref{fig:avJA}, somewhat higher than the empirical values of (7$\pm$2)$\hbar$~\cite{Wilhelmy:1972}. Those empirical values are model-dependent and very sensitive to assumptions made in the nuclear structure and decay probabilities used in the calculations, as discussed in detail in~\cite{Stetcu:2013}.

% !TEX root = ../CGMF-CPC.tex
%-- PHYSICS MODELS: Kinematics of Neutron Emission -----------------------------------------------------
\subsection{Kinematics of neutron emission}
\label{sec:kinematics-neutrons}

We consider here the emission of a neutron from a moving compound nucleus, and assume that the initial momentum of the compound system that is decaying is $\vec P_c$. In the particular case in which the nucleus emitting neutrons is the compound that will undergo fission, we consider first that the incoming neutron with energy $E_n$ is moving along the $z$ axis, so that the neutron momentum $\vec p_n$ is defined as

\begin{equation}
\vec p_n =\sqrt{2M_nE_n}\vec e_z,
\label{eq:p_n}
\end{equation}
where $\vec e_z$ is the unit vector defining the $z$ axis, and $M_n$ is the neutron mass. Using momentum conservation,  $\vec P_c=\vec p_n$, while the velocity of the compound system is then given by
\begin{equation}
\vec V_c=\frac{1}{M_c}\vec P_c,
\end{equation}
with $M_c$ the mass of the compound system. 

We assume that the neutrons are emitted isotropically in the center-of-mass (CM) system (except for pre-equilibrium neutrons which are discussed in detail in Sec. \ref{sec:prefission}), and because we are interested in lab quantities, we will perform transformations between the CM and the lab systems, with all the CM quantities denoted by a prime. Thus, if one neutron is emitted from the compound system, in the lab we have the following momentum conservation law:

\begin{equation}
\vec P_c=\vec P_{c-1}+\vec p_1,
\label{eq:mom_cons1}
\end{equation}
where $\vec p_1$ is the momentum of the first neutron emitted, and $\vec P_{c-1}$ is the momentum of the compound system with one neutron less. Assuming that the energy of the emitted neutron in the center of mass is $\varepsilon_1'$ (previously sampled from the corresponding neutron spectrum), the magnitude of the pre-fission neutron momentum in the CM is determined simply as $p_1'=\sqrt{2\varepsilon_1'M_n}$, while the directions are sampled from an isotropic distribution, which defines the momentum vector $\vec p_1\!'$. In the CM we could determine the recoil of the $c-1$ system $\vec P_{c-1}\!'=-\vec p_1\!'$, and then perform a Galilean transformation for both the pre-fission neutron and the compound system back to the lab system. Alternatively, we transform only the momentum of the pre-fission neutron to the lab system
\begin{equation}
\vec p_1 = \vec p_1\!'+M_n \vec V_c,
\label{eq:mom_cons2}
\end{equation}
and determine $\vec P_{c-1}$ from Eq. (\ref{eq:mom_cons1}). If $n$ neutrons are emitted from the compound system, we iterate Eqs. (\ref{eq:mom_cons1}) and (\ref{eq:mom_cons2}) until all the momenta $(\vec P_{c-2},\vec p_2), \dots, (\vec P_{c-n},\vec p_n)$ are determined, with the input from previous iteration.  Note that here we have discussed only the kinematics, and at this point the neutron multiplicity $n$ is fixed and the corresponding neutron energies $\varepsilon_1\!',\dots,\varepsilon_n\!'$ have been previously sampled. 

The same kinematic transformations can be applied to obtain the momenta of the neutrons emitted from fission fragments. In this case though, the momentum of the incoming nucleon $\vec p_n$ in Eq. (\ref{eq:p_n}) should be replaced by the momentum of the moving fission fragment, which is discussed in the next section. 

%\textcolor{blue}{This section only specifically discusses neutrons emitted before fission; the kinematics are the same though when the neutrons are emitted from the compound.  Should we include that point more explicitly?}

\subsection{Kinematics of the fission fragments}

We now consider the general case of a neutron-induced fission event, in which $n$ pre-fission neutrons are emitted before the $A_c-n$ system fissions. The goal of this section is to determine the momenta of the fission fragments before neutron emission. In the lab system, the previously sampled $TKE$ is given by
\begin{equation}
TKE = \frac{1}{2}M_Lv_L^2+\frac{1}{2}M_Hv_H^2=\frac{1}{2}M_L(\vec v_L\!'+\vec V_{c-n})^2+\frac{1}{2}M_H(\vec v_H\!'+\vec V_{c-n})^2,
\end{equation}
where indices $L$ and $H$ stand for the light and heavy fragment, respectively, and $\vec V_{c-n}=\vec P_{c-n}/M_{c-n}$ is the velocity of the fissioning system after the emission of $n$ neutrons discussed in Sec. \ref{sec:kinematics-neutrons} (note that the formalism works also in the case when $n=0$), with $M_{c-n}$ the mass of the fissioning nucleus. Taking into account the momentum conservation in the CM, $M_L\vec v_L\!'+M_H\vec v_H\!'=\vec 0$, one obtains using simple algebra the connection between $TKE$ in the lab and in the CM respectively

\begin{equation}
TKE=TKE'+\frac{1}{2}(M_L+M_H)V_{c-n}^2.
\label{eq:tke_lab_cm}
\end{equation}
On the other hand, in the CM system of the fissioning system, the magnitude of the fission fragment momenta is given by
\begin{equation}
p_L'=p_H'=\sqrt{2\, TKE' \left(\frac{1}{M_L}+\frac{1}{M_H}\right)^{-1}}.
\end{equation}
To determine the full vectors $\vec p_L\!'$ and $\vec p_H\!'$, the angles, $\theta$ and $\phi$, of the light fragment are sampled, and the direction of the heavy fragment in the CM system is constructed from conservation of momentum.  For most isotopes, both $\mathrm{cos}\theta$ and $\phi$ are sampled isotropically; for $^{235}$U, $^{238}$U, and $^{239}$Pu, $\mathrm{cos}\theta$ is sampled from a $\mathrm{cos}^2\theta$ distribution, discussed further in Sec. \ref{sec:ffad}.  Finally, the FF momenta can be determined simply in the lab frame by performing a Galilean transformation to the lab system, \textit{i.e.},
\begin{equation}
\vec p_L=\vec p_L\!'+M_L\vec V_{c-n},
\end{equation}
and similarly for the heavy fragment (or from momentum conservation $\vec p_H=\vec P_{c-n}-\vec p_L$).

% !TEX root = ../CGMF-CPC.tex

%-- Fission Fragment Angular Distributions
\subsection{Fission Fragment Angular Distributions}
\label{sec:ffad}

Except for spontaneous fission and low-energy neutron-induced fission reactions, the fission fragments are not emitted isotropically in the laboratory frame. At least two processes are responsible for producing an anisotropy. First, the recoil of the compound nucleus when a high-energy neutron hits the target nucleus at rest. This effect is relatively small, even for incident neutron energies up to 20 MeV, but it can become important when studying angular effects on the emission of the neutrons for instance~\cite{Lovell:2020}.

%-- FIGURE: kinematic recoil due to the incident particle impinging on the target
\begin{figure}[ht]
\centerline{\includegraphics[width=0.75\columnwidth]{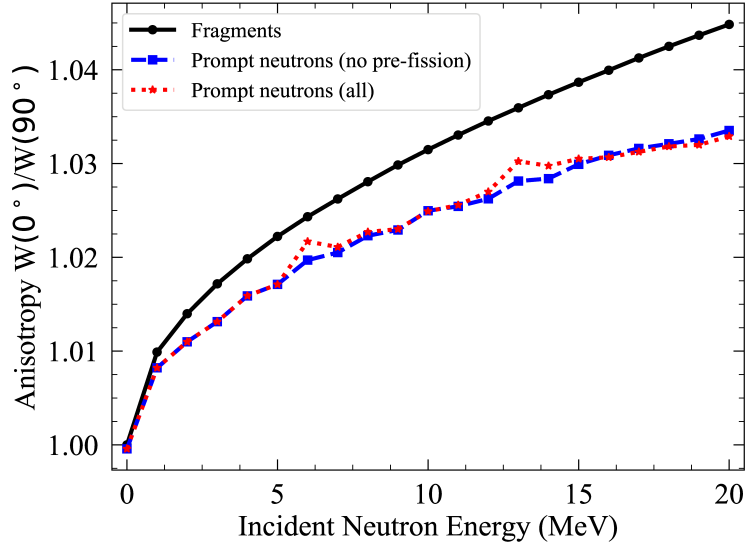}}
\caption{\label{fig:recoil}The kinematic recoil of the fissioning system due to the incoming neutron can lead up to about 5\% anisotropy for the fission fragments in the laboratory frame at 20 MeV incident energy.}
\end{figure}

A second process was identified many decades ago by studying the angular distributions of fission fragments produced in photo-fission reactions. A.~Bohr interpreted~\cite{Bohr:1956} the observed anisotropies in terms of the existence of discrete excited states on top of the outer saddle fission barrier, now called ``fission transition states". Those transition states are characterized by the quantum numbers $(J,K,M)$, where $J$ is the total angular momentum of the compound fissioning nucleus, $K$ its projection on the fission axis, and $M$ its projection on the beam axis. If fission were to occur through only one of those states, the fission fragment angular distribution in the laboratory frame would be characterized by
\begin{eqnarray}
\frac{d\sigma_f}{d\Omega}(J,K,M) = W_{M,K}^J(\theta) = \frac{2J+1}{2} \left| d_{M,K}^J(\theta) \right|^2,
\end{eqnarray}
with
\begin{eqnarray}
\int_0^\pi{W_{M,K}^J(\theta)\sin \theta d\theta}=1.
\end{eqnarray}
The terms $d$ are called the small Wigner $d$-matrices and are given by
\begin{eqnarray}
d^J_{M,K}(\theta)=\sum_n{ (-)^n \frac{\left[ (J+M)!(J-M)!(J+K)!(J-K)! \right]^{1/2}}{(J-M-n)!(J+K-n)!(M-K+n)!n!}
	\cdot \left( \cos \frac{\theta}{2} \right)^{2J+K-M-2n}   \cdot \left( \sin \frac{\theta}{2} \right)^{2n+M-K}     }. 
\end{eqnarray}
The total angular distribution of the fission fragments can then be reconstructed by weighting those matrices according to the population of each transition state:
\begin{eqnarray}
\frac{d\sigma_f}{d\theta} = \sum_s{P_s(J,K,M) \times W_{M,K}^J(\theta)},
\end{eqnarray}
with $P_s$ the population of the transition state $s$ characterized by the quantum numbers ($J$,$K$,$M$).

%-- FIGURE: angular anisotropy for U-235(n,f)
\begin{figure}[ht]
%\centerline{\includegraphics[width=0.75\columnwidth]{figs/ffad-U235.pdf}}
\centerline{\includegraphics[width=0.75\columnwidth]{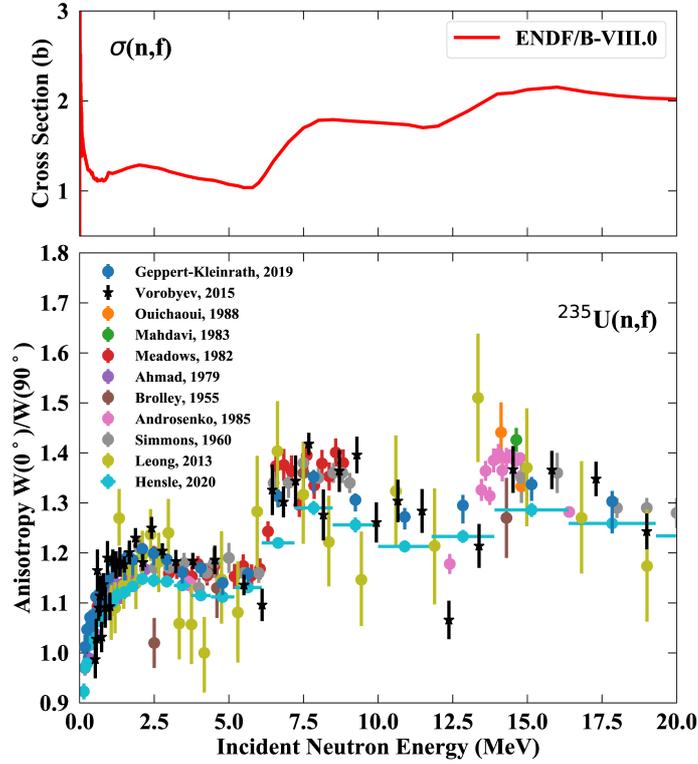}}
\caption{\label{fig:U235-FFAD} (Bottom) The measured anisotropy W(0$^\circ$)/W(90$^\circ$) of the fission fragment angular distribution for the neutron-induced fission reaction on $^{235}$U is shown as a function of incident neutron energy. (Top) The corresponding evaluated fission cross section taken from the ENDF/B-VIII.0 library.}
\end{figure}

The angular distribution of the fission fragments in the neutron-induced fission reaction on $^{235}$U was recently measured at Los Alamos~\cite{GeppertKleinrath2017,Hensle2020} and n\_ToF, CERN~\cite{Leong:2013}. Figure~\ref{fig:U235-FFAD} shows the anisotropy W(0$^\circ$)/W(90$^\circ$) of the fission fragments (bottom) along with the fission cross section (top). 

For very low incident neutron energies, only $s$-wave neutrons contribute and the fragments are emitted isotropically. With increasing excitation energy, the nucleus will fission through specific discrete collective transition states and the anisotropy of the fragments can be understood in terms of the $(J,K,M)$ quantum numbers of those states. At even higher excitation energies, but below the second-chance fission threshold near 6 MeV, more and more transition states are populated, hence smoothing out any specific state anisotropy. Past the second-chance fission threshold, the $(n,n'f)$ process starts populating a few collective states again in the residual $^{235}$U nucleus, hence increasing the anisotropy once more. This process repeats itself for additional multi-chance fission thresholds, but with less intensity each time as more and more multi-chance fission components contribute. Eventually, at very high energies, the fragments are once again expected to be emitted isotropically.

As the prompt fission neutrons are strongly focused along the direction of the fission axis, taking into account this anisotropy is very important for the interpretation of neutron angular distributions at higher incident neutron energies. 

This anisotropy in the angular distributions of fission fragments is taken into account in \CGMF\ for neutron-induced fission reactions on all isotopes considered in \CGMF, where there is somewhat abundant and consistent data on this quantity (see Section~\ref{sec:data} for more details).  Instead of the fission fragment angle being sampled isotropically, the fragments are sampled from a $\mathrm{cos}^2\theta$ distribution,
\begin{equation}
P(\theta) \propto (1-A)\mathrm{cos}^2\theta,
\label{eq:Aprobdist}
\end{equation}
where $A=W(0^\circ)/W(90^\circ)$ is the anisotropy coefficient.  The probability distribution $P(\theta)$ is normalized such that $\int {P(\theta)d\theta}=1$.

%-- FIGURE: 

%\textcolor{red}{
%\begin{itemize}
%\item Add this process in the current version of \CGMF: tabulate LSQ fit of anisotropies for 235U and 239Pu. Others? Implement sampling routine in \CGMF.
%\end{itemize}
%}

% !TEX root = ../CGMF-CPC.tex

%-- PHYSICS MODELS: Pre-Fission Neutron Emission -------------------------------------------------------------
\subsection{Pre-Fission Neutron Emission}
\label{sec:prefission}

Neutrons can be emitted prior to the full acceleration of the fission fragments through several mechanisms. {\it Pre-scission} neutrons can be emitted during the descent from the saddle to the scission point by statistical evaporation. So-called {\it scission} neutrons can also be emitted dynamically at the time of scission, in a process similar to $\alpha$-ternary fission~\cite{Carjan:2007a}. Both processes strongly depend on the dynamical properties of the fission process, which remain largely unknown. Various studies have shown a large spread of estimates for the scission neutron component, from 0 up to 40\% of the total number of emitted neutrons~\cite{Capote:2016}. The time it takes for the fissioning nucleus to go from saddle to scission can strongly influence the number of pre-scission neutrons, which should be reflected in the Coulomb repulsion of the two nascent fragments. 

Another source of pre-scission neutrons appears at higher incident neutron energies, when the multi-chance fission processes become energetically allowed. In that scenario, one or more neutron(s) can be emitted before the residual nucleus fissions. Those reactions are denoted by $(n,n'f)$, $(n,2nf)$, $(n,3nf)$, etc, and called second-chance, third-chance and fourth-chance fission, respectively. At even higher energies ($E_{\rm inc}>20$ MeV), light charged-particle emissions can occur in a similar fashion. In its current version (ver.~\version), \CGMF\ is applicable only to neutron-induced reactions up to 20 MeV incident energy, and pre-fission emissions of light-charge particles are not included.

The multi-chance fission emissions can proceed through an evaporation or a pre-equilbrium phase. In the first scenario, the incident neutron and target nucleus first form a compound nucleus, which then emits one or more neutron(s) in a statistical evaporation process. In the second scenario, a pre-equilibrium reaction occurs before any compound nucleus can form. In both cases, neutrons are emitted prior to fission, but still leaving the residual nucleus with enough excitation energy to undergo fission. However, the neutrons emitted from a compound nucleus evaporation process have a different energy spectrum and angular distribution from the ones emitted in a pre-equilibrium reaction, which tend to be forward-peaked in the laboratory frame. The pre-equilibrium component becomes important only at incident neutron energies above 10 MeV or so.

\begin{figure}[ht]
\centerline{
\includegraphics[width=0.85\columnwidth]{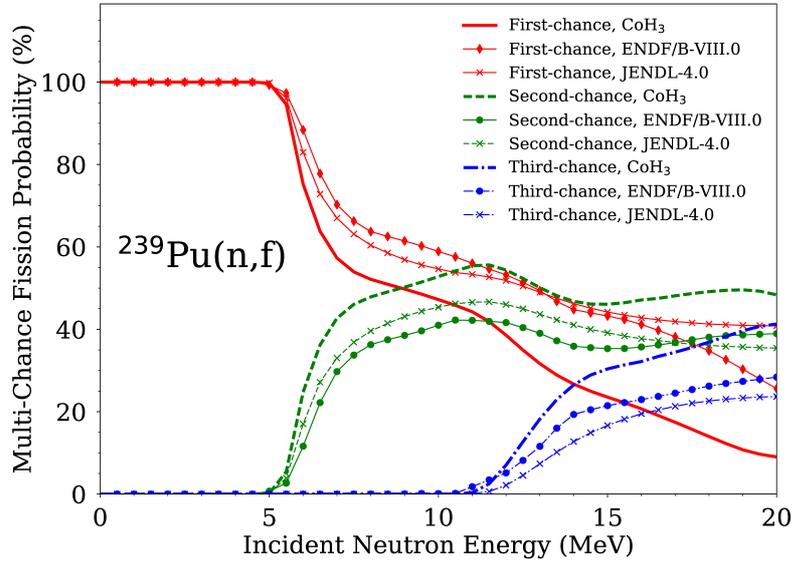}
}
\caption{\label{fig:multichance}First-, second-, and third-chance fission probabilities calculated for the neutron-induced fission reaction on $^{239}$Pu up to 20 MeV by the \COH\ code~\cite{CoH}, and compared to evaluated results from ENDF/B-VIII.0 and JENDL-4.0 libraries.}
\end{figure}

The multi-chance fission probabilities are defined as follow:
\begin{eqnarray}
P_f^i=\frac{\sigma(n,x_inf)}{\sigma(n,f)},
\end{eqnarray}
with $x_i=i$ the number of neutrons emitted prior to fission, and $\sigma(n,f)$ the total neutron-induced fission cross section. These probabilities can be computed from the $\Gamma_n/\Gamma_f$ ratio as a function of the incident neutron energy. This ratio depends in turn on the fission barrier heights in the various compound nuclei $A$, $A-1$, $A-2$, etc. The \COH\ code~\cite{CoH} was used to calculate those ratios for different actinides. Note that both \CGMF\ and \COH\ codes share similar physics models and input parameters, hence this procedure does not introduce any consistency issue. Figure~\ref{fig:multichance} shows the first, second and third-chance fission probabilities in the case of $n+^{239}$Pu up to 20 MeV, in comparison with the ENDF/B-VIII.0 and JENDL-4.0 evaluations. The \COH\ calculations tend to predict a much higher second-chance fission probability at the expense of the first-chance, compared to the evaluations. These quantities are not observables though, and it is therefore difficult to judge the validity of those curves at this point.

%-- FIGURE: multi-chance fission probabilities for 239Pu (n,f)

In \CGMF, those multi-chance fission probabilities are sampled to determine the number of pre-fission neutrons. Then, the energies of those neutrons are obtained by sampling the corresponding neutron spectra. In the case of the first emitted neutron, we keep track of whether the neutron is a pre-equilibrium neutron or evaporated from the compound system and sample directly from either the pre-equilibrium or evaporation spectrum. 

\begin{figure}[ht]
\centerline{
\includegraphics[width=0.9\columnwidth]{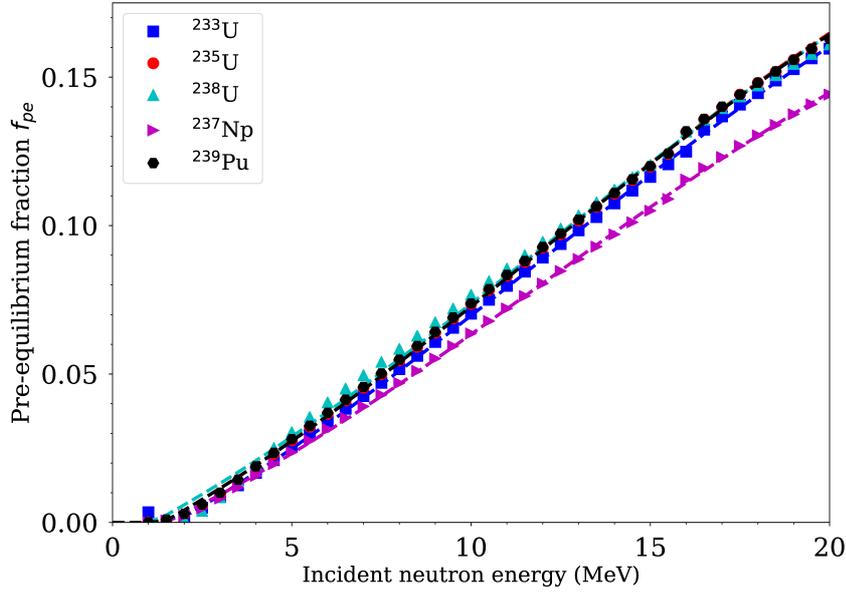}
}
\caption{\label{fig:preeq}Probability of emitting a pre-equilibrium neutron as a function of the incident neutron energy for targets, as calculated with the \COH\ code (symbols), compared against the fit in Eq. (\ref{eq:freeq-fit}) denoted by dashed lines.}
\end{figure}

The fraction of pre-equilibrium neutrons, $f_{pe}$, is also calculated with \COH\ using the exciton model. For each reaction implemented in \CGMF\, the \COH\ results obtained for a set of discrete incident energy points are then used to infer a simple continuous energy-dependent function as follows:
\begin{eqnarray}
f_{pe}(E_{inc}) = \frac{1}{1+\exp\left[ (E_0-E_{inc})/\delta E \right]}+s_E E_{inc} +f_0,
\label{eq:freeq-fit}
\end{eqnarray}
where the parameters $E_0$, $\delta E$, $s_E$ and $f_0$ are fit to \COH\ model calculations.   %given in Table \ref{tab:preeq-params}.
In Fig.~\ref{fig:preeq}, we compare the parameterization of the pre-equilibrium fraction in Eq. (\ref{eq:freeq-fit}) against \COH\ model calculations for a selected set of neutron-induced reactions.

%Pre-equilibrium fractions calculated with the \COH\ code. There is only a slight dependence on the target nucleus, and the fit formula (solid line) is used by default in \CGMF\ instead.

%\begin{table}[h]
%	\centering
%\caption{Parameters for the pre-equlibrium fractions for all reactions currently implemented in \CGMF . \label{tab:preeq-params} }
%\begin{tabular}{cccccc}
%\hline \hline
%Target &  $E_0$ (MeV) & $\delta E$ (MeV) & 100 $\times s_E$ (MeV$^{-1}$) & $ f_0 $ \\
%\hline
%$^{233}$U & 11.913 & 12.948 & -0.97461 & -0.296  \\
%$^{234}$U & 11.913 & 12.948 & -0.91679 & -0.295  \\
%$^{235}$U & 11.913 & 12.948 & -0.96218 & -0.294  \\
%$^{238}$U & 11.913 & 12.948 & -0.97463 & -0.292  \\
%$^{237}$Np & 11.913 & 12.948 & -1.0675 & -0.293 \\
%$^{239}$Pu & 11.913 & 12.948 & -0.96401 & -0.294 \\
%$^{241}$Pu & 11.913 & 12.948 & -0.95544 & -0.293 \\
%\hline\hline
%\end{tabular}
%\end{table}

\begin{figure}[ht]
\centerline{\includegraphics[width=0.95\columnwidth]{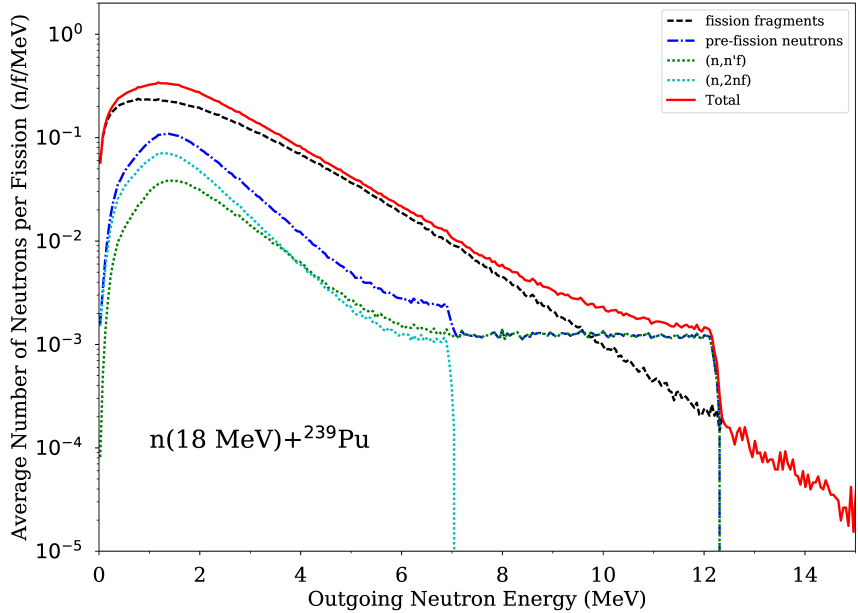}}
\caption{\label{fig:pfns18MeV} Prompt fission neutron spectrum for 18-MeV incident neutrons on $^{239}$Pu. The individual contributions of pre-fission neutrons emitted in a second-chance $(n,n'f)$ or third-chance $(n,2nf)$ fission process, as well as the ones emitted from the fully accelerated fission fragments are shown.}
\end{figure}

Figure~\ref{fig:pfns18MeV} shows the total prompt fission neutron spectrum calculated for 18-MeV incident neutrons on $^{239}$Pu, and distinguishing the individual contributions from all pre-fission neutrons, second-chance $(n,n'f)$, third-chance $(n,2nf)$, and fission fragments.

%-- Figure: PFNS at 18 MeV

The neutrons that are emitted from the compound nucleus have an isotropic angular distribution.  However, the pre-equilibrium neutrons are sampled directly from the forward peaked distribution.  Angular distributions from the model of Feshbach, Kerman, and Koonin (FKK) \cite{Feshbach1980} were calculated for neutron-induced fission on $^{239}$Pu for a two-dimensional grid of incident neutron energies and target excitation energies.  Each angular distribution is described as
\begin{equation}
\frac{d\theta}{d\Omega}(E_n,E^*) = a(E_n,E^*)\mathrm{sinh}(\mathrm{cos}\theta) + b(E_n,E^*),
\label{eq:preeqdist}
\end{equation}
similar to Kalbach systematics \cite{Kalbach1988}.  From the FKK calculations, $a(E_n,E^*)$ and $b(E_n,E^*)$ were globally fit as a function of the incident neutron energy and the target excitation energy, polynomially in both $E_n$ and $E^*$.  For each pre-equilibrium neutron emitted, $\mathrm{cos}\theta$ is sampled from Eq. (\ref{eq:preeqdist}) (normalized to be a probability distribution).  It is assumed that this same angular distribution is applicable for the pre-equilibrium neutrons emitted from any compound nucleus.

% !TEX root = ../CGMF-CPC.tex
%-- PHYSICS MODELS: Hauser-Feshbach ---------------------------------------------------------------------------- 
\subsection{Hauser-Feshbach Statistical Theory of Nuclear Reactions}
\label{sec:HF}

The Hauser-Feshbach statistical theory of nuclear reactions~\cite{Hauser:1952} has been used very successfully over the years to describe a wide range of intermediate energy (keVs to tens of MeV) reactions between light particles and medium to heavy-mass target nuclei. Its simplest representation takes the form of a ratio of decay widths or transmission coefficients to describe the cross section from an entrance channel $a$ and an outgoing channel $b$,
\begin{eqnarray} \label{eq:HF}
\sigma_{ab} \propto T_a\times \frac{T_b}{\sum_c{T_c}},
\end{eqnarray}
in which the sum runs over all possible outgoing channels $c$. The \CGMF\ code implements the Hauser-Feshbach model solely to describe the decay of the fission fragments, and not their initial formation, which would be represented by the entrance channel $a$. One implicit assumption in the derivation of~(\ref{eq:HF}) is that entrance and outgoing channels do not interfere. In addition, one considers the fission fragments to be fully-equilibrated compound nuclei, which is certainly a very good assumption when the fragments are fully separated and accelerated. The present version of the \CGMF\ code does not consider the possibility of scission neutrons, although it handles the emission of pre-fission neutrons from the multi-chance fission process and pre-equilibrium emissions. It also does not consider the case of $(n,\gamma f)$ reactions as studied recently in~\cite{Lynn:2018}.

In the case of low-energy fission fragment de-excitation, only the neutron and \gray\ outgoing channels are considered. This is valid below 20 MeV incident neutron energy, as the probability of emitting charged particles is strongly hindered due to the Coulomb barrier. At higher energies, proton and other light-charged particle emission channels become important~\cite{Tudora:2004}.

% !TEX root = ../CGMF-CPC.tex
%-- PHYSICS MODELS: Neutron Emissions ---------------------------------------------------------------------------- 
\subsection{Neutron Emissions}
\label{sec:neutrons}

Neutron transmission coefficients $T_n^{lj}(\epsilon)$ are obtained through optical model calculations. In this model, the Schr\"odinger equation describing the interaction of incoming waves with a complex mean-field potential is solved, providing the total, shape elastic and reaction cross-sections. It also provides the transmission coefficients that are used in the compound nucleus evaporation calculations.

The transmission coefficients for a channel $c$ are obtained from the scattering matrix $S$ as
\begin{eqnarray} \label{eq:Tn}
T_c=1-\left|\langle S_{cc}\rangle \right|^2.
\end{eqnarray}

To calculate the neutron transmission coefficients for fission fragments, it is important to rely on a global optical model potential (OMP) that can provide results for all fragments that are produced. By default, \CGMF\ uses the global spherical OMP of Koning and Delaroche~\cite{Koning:2003}.

It is important to note that the calculated spectrum of prompt neutrons does depend on the choice of the optical potential used to compute the neutron transmission coefficients. The OMP of Koning-Delaroche has been established to describe a host of experimental data quite well, e.g., total cross-sections, $S_0$ and $S_1$ strength functions, etc. However, those data are only available for nuclei near the valley of stability. Some experimental information do indicate that this optical potential may not be very suitable to the fission fragment region, and therefore a relatively large source of uncertainty in the calculation of the neutron spectrum results from this still open question.

% !TEX root = ../CGMF-CPC.tex
%-- PHYSICS MODELS: Gamma-Ray Emissions --------------------------------------------------------------------------
\subsection{\gray~Emissions}
\label{sec:gammas}

The \gray\ transmission coefficients are obtained using the strength function formalism from the expression: 
\begin{eqnarray}
T^{Xl}(\epsilon_\gamma) = 2\pi f_{Xl}(\epsilon_\gamma)\epsilon_\gamma^{2l+1},
\end{eqnarray}
where $\epsilon_\gamma$ is the energy of the emitted \g\ ray, $X_l$ is the multipolarity of the gamma ray, and $f(X_l)(\epsilon_\gamma)$ is the energy-dependent \gray\ strength function.

For $E1$ transitions, the Kopecky-Uhl~\cite{Kopecky:1990} generalized Lorentzian form for the strength function is used:
\begin{eqnarray}\label{eq:E1SF}
f_{E1}(\epsilon_\gamma,T) = K_{E1}\left[ \frac{\epsilon_\gamma \Gamma_{E1}(\epsilon_\gamma)}{\left( \epsilon_\gamma^2-E_{E1}^2\right)^2 + \epsilon^2_\gamma\Gamma_{E1}(\epsilon_\gamma)^2} +\frac{0.7\Gamma_{E1}4\pi^2T^2}{E_{E1}^5} \right] \sigma_{E1}\Gamma_{E1}
\end{eqnarray}
where $\sigma_{E1}$, $\Gamma_{E1}$, and $E_{E1}$ are the standard giant dipole resonance (GDR) parameters. $\Gamma_{E1}(\epsilon_\gamma)$ is an energy-dependent damping width given by
\begin{eqnarray}
\Gamma_{E1}(\epsilon_\gamma) = \Gamma\frac{\epsilon_\gamma^2+4\pi^2T^2}{E_{E1}^2},
\end{eqnarray}
and $T$ is the nuclear temperature given by
\begin{eqnarray}
T=\sqrt{\frac{E^*-\epsilon_\gamma}{a(S_n)}}.
\end{eqnarray}

The quantity $S_n$ is the neutron separation energy, $E^*$ is the excitation energy of the nucleus, and $a$ is the level density parameter. The quantity $K_{E1}$ is obtained from normalization to experimental data on $2\pi\langle \Gamma_{\gamma_0} \rangle / \langle D_0 \rangle$. 

For $E2$ and $M1$ transitions, the Brink-Axel~\cite{Brink:1957,Axel:1962} standard Lorentzian is used instead:
\begin{eqnarray} \label{eq:E2M1SF}
f_{Xl}(\epsilon_\gamma)=K_{Xl}\frac{\sigma_{Xl}\epsilon_\gamma\Gamma_{Xl}^2}{(\epsilon_\gamma^2-E_{Xl}^2)^2+\epsilon_\gamma^2\Gamma_{Xl}^2}.
\end{eqnarray}
In the current version of \CGMF\ (ver.~\version), only $E1$, $E2$, and $M1$ transitions are allowed, and higher multipolarity transitions are neglected.

% !TEX root = ../CGMF-CPC.tex

%-- PHYSICS MODELS: Low-Lying Nuclear Structure  -------------------------------------------------------------
\subsection{Nuclear Structure and Level Density}
\label{sec:structure}

Knowing the nuclear structure of the neutron-rich nuclei produced in the fission process is very important to compute the final \g\ decay chain accurately. Fission yields, as well as isomeric ratios, are often inferred from the measurement of specific \g\ lines, often using \g-\g\ or \g-\g-\g\ coincidences to clean up the complicated observed \g\ spectrum.

In \CGMF, discrete levels are initially obtained from the RIPL-3 database, but special treatment is necessary when only limited or incomplete information is given. This modification of the nuclear structure data used in \CGMF\ is done prior to its use in the code. The complete set of possible modifications of the original data file is as follows:

\begin{itemize}
\item If RIPL provides a set of possible spins, the first spin specified -- usually the lowest one -- is chosen.
\item The default spin/parity values for the ground state is: $0^+$ for even-even, $1^+$ for odd-odd, and $1/2^+$ for even-odd fragments, respectively.
\item For higher-energy states, default spin/parity values will be selected from a simple Gaussian distribution, with a spin cut-off parameter chosen according to the Kawano-Chiba-Koura (KCK) level density systematics~\cite{Kawano:2006}. The parity is chosen randomly.
\item For discrete excited states included in the nuclear structure but for which no decay is reported, it is assumed that 100\% of the decay will be through \g\ lines of E1 nature. The strength of the decay will be distributed equally among levels lying below this particular excited state.
\end{itemize}

\begin{figure}[ht]
\centerline{\includegraphics[width=0.75\columnwidth]{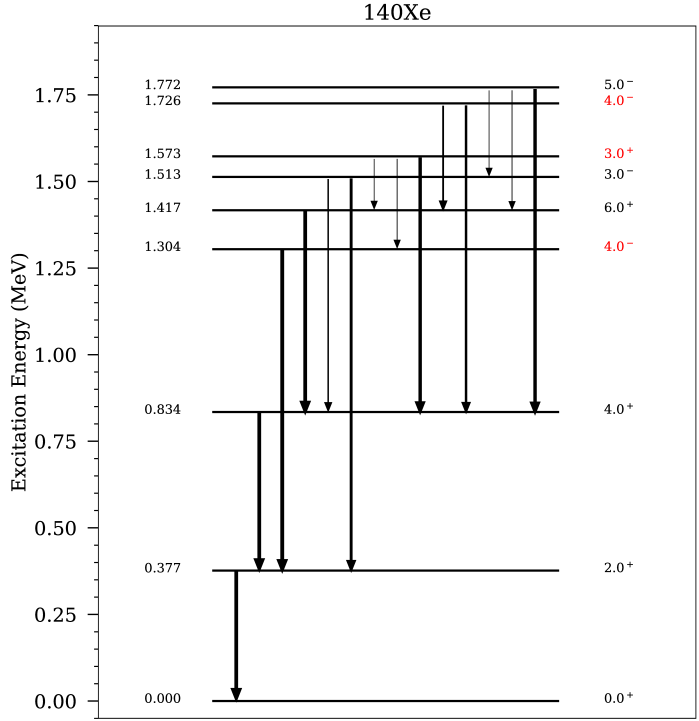}}
\caption{\label{fig:140Xe} Example of a default augmentation of the incomplete nuclear structure for $^{140}$Xe obtained from the RIPL-3 database~\cite{RIPL3} for use in \CGMF. In this case, only the spin and parity of 3 states were modified (in red).}
\end{figure}

An example of such modifications is shown in Fig.~\ref{fig:140Xe} in the case of $^{140}$Xe.

%-- FIGURE: 140Xe 

Many of these assumptions are rather arbitrary. When studying a specific reaction, fission yield or \g\ spectrum for a specific fragment, those default assumptions should be revisited in light of a more careful analysis of available experimental and theoretical information.

% !TEX root = ../CGMF-CPC.tex

%-- PHYSICS MODELS: Continuum Level Density ------------------------------------------------------
\subsection{Continuum Level Density}
\label{sec:leveldensity}

In \CGMF, the Gilbert-Cameron~\cite{Gilbert:1965} model of level densities is used for all fragments. In this model, a constant-temperature formula is used to represent the level density at lower excitation energies, while a Fermi-gas formula is used at higher excitation energies. Experimental data on the average level spacing at the neutron separation energy can be used to constrain parameters entering the Fermi-gas formula, while low-lying discrete levels are used to constrain the constant-temperature parameters. Again, little data is available for nuclei far from stability where systematics have been developed, contributing to uncertainties in the final predicted data. 

\begin{figure}[ht]
\centerline{\includegraphics[width=0.85\columnwidth]{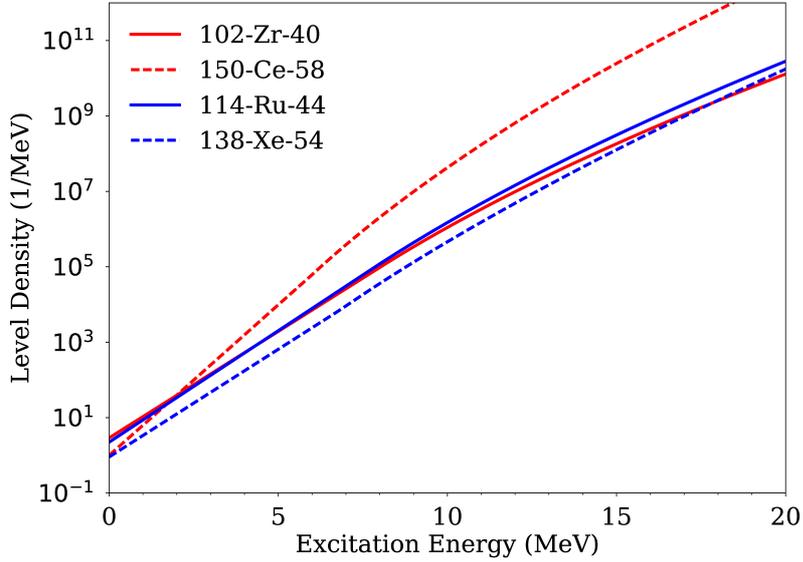}}
\caption{\label{fig:levelDensity}Level densities for two pairs of complementary fission fragments are plotted as a function of excitation energy. They are obtained using Eqs.~\ref{eq:CT} and~\ref{eq:FG} for the constant-temperature and Fermi-gas regions at low and high excitation energies, respectively.}
\end{figure}

The constant-temperature form is given by
\begin{eqnarray} \label{eq:CT}
\rho_{CT}(U)=\frac{1}{T}{\rm exp}\left( \frac{U+\Delta-E_0}{T} \right),
\end{eqnarray}
where $T$ is the nuclear temperature and $E_0$ is a normalization factor. The quantity $U$ is the excitation energy $E$ minus the pairing energy $\Delta$. At higher excitation energies, the Fermi-gas form of the level density is used instead and is given by

\begin{eqnarray} \label{eq:FG}
\rho_{FG}(U)=\frac{{\rm exp}\left( 2\sqrt{aU}\right)}{12\sqrt{2}\sigma(U)U(aU)^{1/4}},
\end{eqnarray}
where $a$ is the level density parameter. The constant-temperature form of the level density is matched to cumulative low-lying discrete levels, when they are known. For fission fragments, which are neutron-rich and rather poorly known, this constant-temperature level density is sometimes used down to the ground-state. Figure~\ref{fig:levelDensity} shows examples of level densities calculated for two pairs of fission fragments.

%-- FIGURE: level density calculations

The energy, spin and parity-dependent level density is then obtained as
\begin{eqnarray} \label{eq:levden_spin}
\rho(U,J,\pi) = \frac{1}{2}\cdot\frac{2J+1}{2\sigma^2}\exp\left\{ -\frac{(J+1/2)^2}{2\sigma^2}\right\}\cdot \rho(U),
\end{eqnarray}
with $\rho(U)$ given by either the constant-temperature (Eq.~\ref{eq:CT}) or Fermi-gas (Eq.~\ref{eq:FG}) formula. This equation assumes equiprobability between positive and negative parity states in the continuum. The quantity $\sigma$ is the spin cut-off parameter fitted to reproduce the observed distribution of low-lying levels.

In its original formulation, the Gilbert-Cameron formalism uses an energy-independent level density parameter $a$. To better describe the washing-out of shell effects at higher excitation energies, Ignatyuk~\cite{Ignatyuk:1975} developed a model that uses an energy functional for the level density parameter as
\begin{eqnarray} \label{eq:Ignatyuk}
a(U) = \tilde{a} \left( 1+\delta W \frac{1-{\rm exp}(-\gamma U)}{U} \right).
\end{eqnarray}

In this formula, $\tilde{a}$ is the asymptotic value of the level density parameter at high energy, $\delta W$ is the shell correction energy, and $\gamma$ is an empirical damping width to account for the washing-out of shell effects at high energy.

In \CGMF, the Kawano-Chiba-Koura level density parameterization~\cite{Kawano:2006} is used. Parameters were updated since the original publication, so we are describing them here. The KTUY05 mass model was used to compute the shell correction and pairing energies. The asymptotical level density parameter $a^*$ is expressed as a function of the mass number $A$,
\begin{eqnarray} \label{eq:levden_asympt}
a^*=0.126181\cdot A + 7.52191\cdot 10^{-5}A^2.
\end{eqnarray}
In this formula, the damping width $\gamma$ was estimated as
\begin{eqnarray}
\gamma=0.31A^{-1/3}.
\end{eqnarray}
The spin cut-off parameter appearing in Eq.~\ref{eq:levden_spin} is given by
\begin{eqnarray}
\sigma^2=0.006945\sqrt{\frac{U}{a}}A^{5/3}.
\end{eqnarray}
At low energies, systematics for the constant-temperature model were also obtained by connecting the cumulative number of levels obtained with the Fermi-gas level density at higher energies and with the available discrete level data at lower energies for more than 1,000 nuclei. Systematics for the parameters $E_0$ and $T$ entering in Eq.~\ref{eq:CT} are
\begin{eqnarray} \label{eq:levden_temperature}
T = 47.0582 \cdot A^{-0.893714} \sqrt{1-0.1 \delta W},
\end{eqnarray}
and 
\begin{eqnarray} \label{eq:levden_E0}
E_0=\Delta-0.16\cdot \delta W+(-0.01380\cdot A-1.379) \cdot T,
\end{eqnarray}
where $\Delta$ and $\delta W$ are the pairing and shell correction energies, respectively, taken from the mass formula of Koura \etal (KTUY05)~\cite{Koura:2005}.

%-- FIGURE: Cumulative levels for different fission fragments
%\begin{figure}[ht]
%\centerline{\includegraphics[width=0.75\columnwidth]{figs/cumulativeLevels.pdf}}
%\caption{\label{fig:cumulative}Cumulative levels as a function of excitation energy for different fission fragments.}
%\end{figure}

		%-- Physics models (subsections inside)
% !TEX root = ../CGMF-CPC.tex
%-- AUXILIARY DATA FILES ----------------------------------------------------------------------------------
\section{Auxiliary Data Files}
\label{sec:data}

Various model input parameters and auxiliary data are needed for \CGMF\ to run. The complete list is provided in Table~\ref{tab:datafiles}. Some brief description of each data file follows.

%-- Table of all CGMF data files
\begin{table}[h]
\def\arraystretch{1.15}
\begin{center}
\caption{List of auxiliary data files (in alphabetical order) needed to run \CGMF.}
\resizebox{\columnwidth}{!} {
\begin{tabular}{ll}
\hline
File name & Description \\
\hline
\texttt{anisotropy.dat} & Fission fragment angular anisotropies as a function of energy \\
\texttt{deformations.dat} & Ground-state deformation properties based on the FRDM model \\
\texttt{discreteLevels.dat} & Experimental database of identified discrete levels in all nuclei \\
%\texttt{gaussYA.dat} & Five-Gaussian mass yield parameters \\
\texttt{kcksyst.dat} & Level density parameters from the Kawano-Chiba-Koura systematics~\cite{Kawano:2006} \\
\texttt{ldp.dat} & Level density parameters for all nuclei as a function of excitation energy \\
\texttt{masses.dat} & Audi 2011 ground-state mass table \\
\texttt{multichancefission.dat} & Energy-dependent multi-chance fission probabilities \\
\texttt{preeq-params.dat} & Energy-dependent pre-equilibrium probabilities \\
\texttt{preeq-scattering-params.dat} & Angular distributions of pre-equilibrium neutrons \\
%\texttt{ripl3-levels.dat} & \textcolor{red}{[All] Is this file used by CGMF?} \\
\texttt{rta.dat} & $R_T(A)$ mass-dependent temperature ratios \\
\texttt{spinscalingmodel.dat} & Fit parameters for the energy-dependence of the spin-scaling factor $\alpha$ \\
\texttt{temperatures.dat} & Energy-dependent nuclear temperatures as a function of ZAID and excitation energy \\
\texttt{tkemodel.dat} & Fit parameters for the total kinetic energy distribution \\
\texttt{yamodel.dat} & Fit parameters for the five-Gaussian mass yield parameters \\
\hline
\end{tabular}
}
\label{tab:datafiles}
\end{center}
\end{table}

%-- anisotropy.dat
\subsection{Fission fragment angular distribution: \texttt{anisotropy.dat}}

\begin{figure}[ht]
\centering
\includegraphics[width=0.75\textwidth]{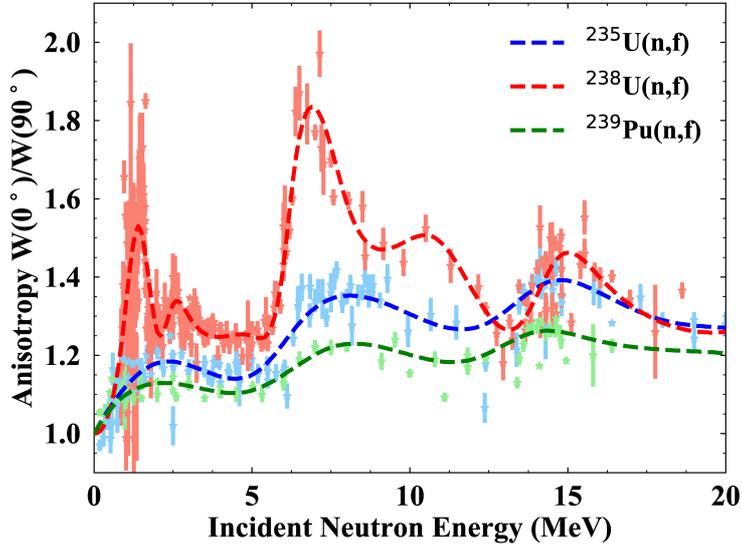}
\caption{Cubic spline fits (dashed lines) to experimental data (points) for the anisotropy of fission fragments from neutron-induced fission of $^{235}$U \cite{Vorobyev2015,Ahmad1979,Brolley1955,Simmons1960,GeppertKleinrath2017}, $^{238}$U \cite{Vorobyev2015,Henkel1956,Brolley1954,Birgersson2009,Ryzhov2005,Vives2000,Simmons1960}, and $^{239}$Pu \cite{Leachman1965,Blumberg1959,Simmons1960} as a function of the incident neutron energy from thermal up to 20 MeV.}
\label{fig:anisotropy}
\end{figure}

This file contains fission fragment anisotropies as a function of incident neutron energy for all neutron-induced reactions handled in {\CGMF}.  Anisotropies are defined as the ratio of cross sections at $0^\circ$ to $90^\circ$.  Available experimental data on this quantity have been fitted with cubic splines and are tabulated from thermal energies up to 20 MeV in steps of 0.1 MeV and are interpolated between these points within {\CGMF}.  The only exception is for $^{241}$Pu where experimental data is only available up to an incident neutron energy of 8 MeV.  In this case, the spline fit for $^{239}$Pu is used.  The results of the fits for $^{235}$U, $^{238}$U, and $^{239}$Pu are shown in Fig.~\ref{fig:anisotropy}.

%-- FIGURE: Anisotropy W(0)/W(90) for the Big Three

%-- deformations.dat
\subsection{Ground-state deformation properties of fission fragments: \texttt{deformations.dat}}

Masses and ground-state deformation properties ($\beta_2$ through $\beta_6$) of the fission fragments are read from the file \texttt{deformations.dat}. It contains results from calculations within the Finite Range Droplet Model (FRDM, 1995) by M\"oller \etal~\cite{Moller:1995}. Details on the content of this file and format can be found in the RIPL-3 documentation~\cite{RIPL3} and on its corresponding website: https://www-nds.iaea.org/RIPL-3/. 

%-- discreteLevels.dat
\subsection{Nuclear structure file: \texttt{discreteLevels.dat}}

Nuclear structure information for all fission fragments produced is obtained from the \RIPL\ library~\cite{RIPL3}, which is itself derived from the Evaluated Nuclear Structure Data File (ENSDF) library, and stored in the \texttt{discreteLevels.dat} file. Modifications to those files are sometimes needed, as explained in Sec.~\ref{sec:structure}. Data are provided for many nuclei, and for each nucleus, the data are provided in the following format:

\texttt{
\begin{itemize}
\item $^A$Symbol - ZAID - $N_{\rm lev}$ \\ \\
\begin{itemize}
\item $E_1$, $J_1$, $\pi_1$, $T_{1/2}^1$, $N_\gamma^1$
\begin{itemize}
\item I$_f^1$, $P_{i\rightarrow f}^1$, ICC$_\gamma^1$
\item ...
\item I$_f^{N_\gamma^1}$, $P_{i\rightarrow f}^{N_\gamma^1}$, ICC$_\gamma^{N_\gamma^1}$
\end{itemize}
\item $E_2$, $J_2$, $\pi_2$, $T_{1/2}^2$, $N_\gamma^2$
\begin{itemize}
\item ...
\end{itemize}
\item ... 
\item $E_{N_{\rm lev}}$, $J_{N_{\rm lev}}$, $\pi_{N_{\rm lev}}$, $T_{1/2}^{N_{\rm lev}}$, $N_\gamma^{N_{\rm lev}}$
\end{itemize}
\end{itemize}
}

As an example, here is the beginning of the data entry for $^{130}$Sn (Z=50), which includes 13 discrete states.

\bigskip
\noindent
\texttt{130Sn  50130  13 \hfill{\rm Z=50, A=130, 13 discrete levels} \\
 0.000000   0.0  1  2.23000e+02  0 \hfill{\rm $E_0$=0.0 MeV, 0$^+$, $T_{1/2}=223$ sec, no decay line}\\
 1.221260   2.0  1  0.00000e+00  1 \hfill{\rm $E_1$=1.22 MeV, $2^+$, stable, one \g\ line} \\
 0 1.000e+00 8.149e-04 \hfill{\rm 100\% decaying to the level \#0 (g.s.), ICC=$8.149\cdot 10^{-4}$} \\
 1.946880   7.0 -1  1.02000e+02  1\\
 1 1.000e+00 0.000e+00\\
 1.995620   4.0  1  0.00000e+00  1\\
 1 1.000e+00 2.269e-03\\
 2.028310   2.0  1  0.00000e+00  2\\
 ...\\
}

%-- kcksyst.dat
\subsection{Level density parameter systematics: \texttt{kcksyst.dat}}

Level density parameters from the Kawano-Chiba-Koura systematics~\cite{Kawano:2006} are saved in the \texttt{kcksyst.dat} file. The following quantities are given for each entry (line):

\bigskip
\noindent
\texttt{Z     A   pairing      Eshell        Ematch    a*        T         E0        T(sys)    E0(sys)}

\bigskip
Each line corresponds to a different nucleus characterized by its charge and mass numbers, given by the first two entries. The pairing $\Delta$ and shell correction $\delta W$ energies, appearing in Eq.~\ref{eq:levden_E0}, are then provided as \texttt{pairing} and \texttt{Eshell}. All energies provided in this file are given in MeV. The matching energy between the constant temperature and Fermi gas regions is given as \texttt{Ematch}. The value of the asymptotic level density parameter given in Eq.~\ref{eq:levden_asympt} is provided by \texttt{a*}. The temperature $T$ and energy $E_0$ appearing in Eqs.~\ref{eq:levden_temperature} and~\ref{eq:levden_E0} are then given as \texttt{T} and \texttt{E0}. 

\bigskip
\noindent
As an example, let us consider the line for $^{144}$Xe:

\bigskip
\noindent
\texttt{\small 54   144  1.96606e+00 -2.34600e+00   0.00000  19.72981   0.00000   0.00000   0.61580   0.26851}

\bigskip
\noindent
which translates into
\begin{itemize}
\item Z=54
\item A=144
\item $\Delta$=1.96606 MeV
\item $\delta W$=-2.346 MeV
\item $E_{\rm match}$=0.0
\item $a^*$= 0.126181$\cdot$ 144 + 7.52191 $\cdot$ 10$^{-5}\cdot$ 144$^2$ = 19.72981 MeV$^{-1}$ {\it (Eq.~\ref{eq:levden_asympt})}
\item $T$=47.0582 $\cdot 144^{-0.893714}\cdot \sqrt{1-0.1\cdot(-2.346)}$ = 0.61580 MeV {\it (Eq.~\ref{eq:levden_temperature})}
\item $E_0$=1.96606-0.16$\cdot$(-2.346) + (-0.01380$\cdot$144-1.379)$\cdot$ 0.61580 =  0.26851 MeV {\it (Eq.~\ref{eq:levden_E0})}
\end{itemize}

%-- ldp.dat
\subsection{Level density parameters: \texttt{ldp.dat}}

Instead of calculating the energy-dependent level density parameter $a(U)$ from Eq.~\ref{eq:Ignatyuk} inline, those values have been calculated and tabulated separately on a given energy grid from 0.0 to 100 MeV for all possible fragments produced in a \CGMF\ run. The tabulated values are given in the file \texttt{ldp.dat}. The first data line provides the energy grid from 0.0 to 100.0 MeV, with steps of 1.0 MeV up to 10.0 MeV, steps of 2.0 MeV from 10.0 to 20.0 MeV, and steps of 5.0 MeV beyond 20.0 MeV. The rest of the file contains the tabulated $a(U_i)$ values for a given \texttt{ZAID} corresponding to a fission fragment.

%-- masses.dat
\subsection{Ground-state nuclear masses: \texttt{masses.dat}}

Nuclear masses are an important ingredient in \CGMF\ calculations as they determine the $Q$ value of the fission reaction for all specific fission fragmentations, as well as the neutron separation energies used along the decay chain of subsequent neutron emissions. Experimental nuclear ground-state masses are taken from the Audi 2012 Atomic Mass Evaluation (AME2012)~\cite{Audi:2012}. For nuclei whose mass has not been measured, theoretically-predicted masses obtained in the Finite-Range Droplet Model (FRDM, 1995)~\cite{Moller:1995} are used instead.

%-- multichancefission.dat
\subsection{Multi-chance fission probabilities: \texttt{multichancefission.dat}}

This file contains the energy-dependent multi-chance fission probabilities as calculated with the \COH\ code for a suite of neutron-induced fission reactions and isotopes. The probabilities are calculated up to the fourth-chance fission up to 20 MeV incident neutron energy.

The file first contains the ZAID number of the target, the number of incident neutron energies at which the probabilities are calculated $\mathtt{N_E}$, and then the maximum number of pre-fission neutrons that can be emitted $\mathtt{N_{PF}}$:
\begin{eqnarray}
\mathtt{ZAIDt}\mbox{ }\mathtt{N_E}\mbox{ }\mathtt{N_{PF}}.
\end{eqnarray}
The next line gives the height (in MeV) of the highest fission barrier, \texttt{E$_f^i$} for the $i^{\mathrm{th}}$-chance fission,
\begin{eqnarray}
\mathtt{E_f^1}\mbox{ }\mathtt{E_f^2}\mbox{ }\mathtt{E_f^3}\mbox{ }\mathtt{E_f^4}.
\end{eqnarray}
The following $\mathtt{N_E}$ lines provide the incident neutron energy and the probabilities for all four chance fissions.

%The file first contains the max, min, and step size of the incident neutron energy grid on which the probabilities have been calculated:
%\begin{eqnarray}
%\mathtt{E_{min}}\mbox{ }\mathtt{E_{max}}\mbox{ }\mathtt{E_{step}},
%\end{eqnarray}
%where all energies are given in MeV. For each target nucleus, the format of the file is as follows:
%\begin{eqnarray}
%\mathtt{ZAIDt}\mbox{ }\mathtt{E_f^1}\mbox{ }\mathtt{E_f^2}\mbox{ }\mathtt{E_f^3}\mbox{ }\mathtt{E_f^4},
%\end{eqnarray}
%where \texttt{ZAIDt} is the ZAID number of the target nucleus, and \texttt{E$_f^i$} is the height (in MeV) of the highest fission barrier for the $i^{\rm th}$-chance fission. The probabilities calculated on the energy grid defined are then provided for all four chance fissions.

%-- Pre-equilibrium fractions
\subsection{Pre-equilibrium fractions: \texttt{preeq-params.dat}}

The fractions of pre-equilibrium neutrons as a function of incident neutron energy were calculated for all target nuclei using the CoH 3.5.3 code~\cite{CoH}, and are shown in Fig.~\ref{fig:preeq}. They are contained in this data file, and follows the following format:
\begin{eqnarray}
\mathtt{ZAIDt}\mbox{ }\mathtt{e0}\mbox{ }\mathtt{de}\mbox{ }\mathtt{se}\mbox{ }\mathtt{f0},
\end{eqnarray}
with $\mathtt{ZAIDt}$ the ZAID of the target nucleus, and ($\mathtt{e0}, \mathtt{de}, \mathtt{se}, \mathtt{f0}$) corresponding to the ($E_0, \delta E, s_E$, $f_0$) parameters entering Eq.~\ref{eq:freeq-fit}, respectively. %The calculated parameter values are summarized in Table~\ref{tab:preeq-params}.

%-- Pre-equilibrium angular distribution parameters
\subsection{Pre-equilibrium angular distribution parameters: \texttt{preeq-scattering-params.dat}}

The angular distribution of the neutrons emitted in the pre-equilibrium process has been calculated using the FKK theory, and then fitted according to
\begin{eqnarray}
p_0\sinh \left( \cos (\theta)\right)+p_1,
\end{eqnarray}
where the parameters $p_0$ and $p_1$ use the same functional form in terms of the incident energy $E_n$ and excitation energy $E^*$:
\begin{eqnarray}
p_i=a_i+b_i E_n+c_i E_n^2+d_i E_n^3,
\end{eqnarray}
with
\begin{eqnarray}
a_i, b_i, c_i, d_i = \sum_{j=0}^{5}{u_j[E^*]^j}.
\end{eqnarray}
The \texttt{preeq-scattering-params.dat} data file contains those parameters. There is a total of 4$\times$6=24 parameters for both $p_0$ and $p_1$. The data file first lists the 24 parameters for $p_0$ in the form of the $u_j$ values for $(a_0,b_0,c_0,d_0)$, followed by a similar list of parameters for $p_1$.

%-- R_T(A)
\subsection{Mass-dependent temperature ratios: \texttt{rta.dat}}

The mass-dependent ratio $R_T$(A) introduced in Eq.~\ref{eq:RT} has been fitted to experimental data on \nubar(A) for all isotopes handled by \CGMF, wherever experimental data exist. The data file \texttt{rta.dat} contains those ratio values following the format:

\bigskip
\noindent
\texttt{ZAIDt Amin Amax $R_T$(Amin:Amax),}

\bigskip
\noindent
where \texttt{ZAIDt} corresponds to the ZAID of the target nucleus, and \texttt{(Amin,Amax)} define the range of heavy fragment mass values for which the ratio is given. As usual, if a negative value is given, it corresponds to the compound nucleus in a spontaneous fission reaction. If no data is given in the file, \CGMF~reverts to a default ratio of 1.2.

%-- Spin-scaling model
\subsection{Spin-scaling model: \texttt{spinscalingmodel.dat}}

The spin-scaling parameterization discussed in Section~\ref{sec:spin} follows the simple linear relation as a function of incident neutron energy $E_n$:
\begin{eqnarray}
\alpha(E_n)=\alpha_0+\alpha_1E_n.
\end{eqnarray}
Those parameters are given in the \texttt{spinscalingmodel.dat} file, following:
\begin{eqnarray}
\mathtt{ZAIDt}\mbox{ }\mathtt{\alpha_0}\mbox{ }\mathtt{\alpha_1}
\end{eqnarray}
As usual, in the case of spontaneous fission, $\mathtt{ZAIDt}$ is set negative. For all uranium and neptunium isotopes, those fits were obtained following Ref.~\cite{Frehaut:1983}.

%-- temperatures.dat
\subsection{Nuclear temperatures: \texttt{temperatures.dat}}

\begin{figure}[ht]
\centerline{\includegraphics[width=0.75\columnwidth]{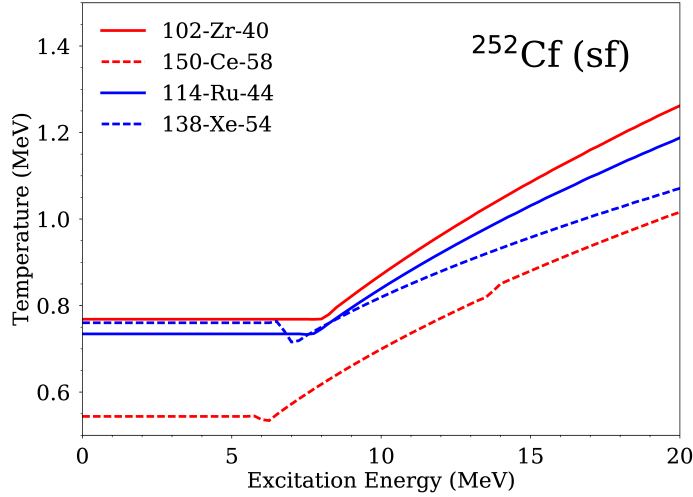}}
\caption{\label{fig:temperatures}Nuclear temperatures calculated using Eq.~\ref{eq:temperature} for two pairs of fission fragments (same as in Fig.~\ref{fig:levelDensity}).}
\end{figure}

Nuclear temperatures are calculated for all nuclei and read in before any \CGMF\ calculation takes place. They are used in the calculation of the sorting of the total excitation energy between the two complementary fragments at scission. For a given nucleus, the temperature $T$ is calculated as:
\begin{eqnarray} \label{eq:temperature}
T^{-1} =  \frac{d} {dU} \log \rho( U ),
\end{eqnarray}
where $\rho$ is the total level density at the excitation energy $U$. Figure~\ref{fig:temperatures} shows an example of energy-dependent nuclear temperatures calculated for a couple of fission pairs. For low excitation energies, we recognize the expected constant-temperature behavior followed by an increase due to the transition to the Fermi gas regime. In general, the light fragment would exhibit a lower temperature than its heavy partner. Exceptions to this rule exist, especially since fission fragments are produced predominantly in shell closure regions where level densities are significantly different than in transitional nuclei.

%-- FIGURE: Temperatures vs. Excitation Energy

%-- tkemodel.dat
\subsection{Fit parameters for the description of the total kinetic energy yields: \texttt{tkemodel.dat}}

The fit parameters describing the energy-dependence of the average TKE and the mass dependence of $\overline{\mathrm{TKE}}$ and $\sigma_\mathrm{TKE}$, as described in Section \ref{sec:yields}, are tabulated in this file.  Each line corresponds to a separate fission reaction:

\bigskip
\noindent
\texttt{ZAIDt  $\kappa^{(0)}$  $E_c$ $\kappa^{(1)}$ $\rho^{(1)}$ $A_0^\mathrm{TKE}$ $A_\mathrm{max}^\mathrm{TKE}$ $a_0$ $a_1$ $a_2$ $a_3$ $a_4$ $a_5$ $a_6$ $a_7$ $a_8$ $A_0^{\sigma_\mathrm{TKE}}$ $A_\mathrm{max}^{\sigma_\mathrm{TKE}}$ $b_0$ $b_1$ $b_2$ $b_3$ $b_4$ $b_5$ $b_6$ $b_7$ $b_8$,}

\bigskip
\noindent providing the parameters for $\overline{\mathrm{TKE}}(E_n)$, $\overline{\mathrm{TKE}}(A_H)$, and $\sigma_\mathrm{TKE}(A_H)$ as defined in Section \ref{sec:yields}.  In Eq. (\ref{eq:TKEE}), $\rho^{(0)}$ is determined by the continuity at $E_c$.  Note that each of these polynomial parameterizations contains a cut-off mass, $A_\mathrm{max}$ beyond which $\mathrm{TKE}(A_H>A_\mathrm{max}^\mathrm{TKE})=140$ MeV and $\sigma_\mathrm{TKE}(A_H>A_\mathrm{max}^{\sigma_\mathrm{TKE}})=7$ MeV.  As with the mass yields, \texttt{ZAIDt} corresponds to the ZAID of the target nucleus, and if \texttt{ZAIDt} is negative, this reaction is the spontaneous fission of the target nucleus.

%-- yamodel.dat
\subsection{Gaussian fit parameters for the description of mass yields: \texttt{yamodel.dat}}

The parameters for the energy-dependent gaussian fits of Y(A;E$_{\rm inc}$), as described in Section~\ref{sec:yields}, are tabulated in the file \texttt{gaussYA.dat}. Each line corresponds to a separate fission reaction:

\bigskip
\noindent
\texttt{ZAIDt  $w_a^{(1)}$  $w_b^{(1)}$ $\mu_a^{(1)}$ $\mu_b^{(1)}$ $\sigma_a^{(1)}$ $\sigma_b^{(1)}$ $w_a^{(2)}$  $w_b^{(2)}$ $\mu_a^{(2)}$ $\mu_b^{(2)}$ $\sigma_a^{(2)}$ $\sigma_b^{(2)}$ $\sigma_a^{(3)}$ $\sigma_b^{(3)}$,}

\bigskip
\noindent
providing the parameters for two asymmetric modes and one symmetric mode, following the definitions established in Section~\ref{sec:yields}. \texttt{ZAIDt} corresponds to the ZAID of the target nucleus. If negative, \texttt{ZAIDt} corresponds to a spontaneous fission reaction, and the ``target" is identical to the compound nucleus.

		%-- Auxiliary data files needed to run the code
% !TEX root = ../CGMF-CPC.tex
%== MCNP-6.2 INTEGRATION ===============================================
\section{MCNP6\textsuperscript{\textregistered} Version 6.2 Code Integration}
\label{sec:mcnp}

A version of \CGMF\, prior to the open-source release, has already been integrated into the \MCNP\ version 6.2 transport code~\cite{TechReport_2018_LANL_LA-UR-18-20808_WernerBullEtAl}, and we expect the present open-source version (\version) of \CGMF\ to be incorporated in the next official release of the \MCNP\ code as well.  While many of the details of the integration, user-interface, and verification testing can be found in a variety of references~\cite{TechReport_2017_LANL_LA-UR-17-29981_WernerArmstrongEtAl,TechReport_2016_LANL_LA-UR-16-23341_Rising,TechReport_2017_LANL_LA-UR-16-27710_Rising,TechReport_2017_LANL_LA-UR-17-20799_RisingSood}, a brief summary is provided here.

By default in the \MCNP\ transport code, the bounded integer sampling scheme is employed to simulate secondary neutrons emitted from fission reactions, relying entirely on the average or expected number of neutrons emitted in fission, $\bar{\nu}(E)$.  When a neutron-inducing fission event occurs for a given target $ZA$ isotope and a given incident neutron energy $E$, the number of neutrons emitted is randomly selected to be either the largest integer number above or below $\bar{\nu}(E)$.  The probabilities of selecting these two integers are chosen to preserve the average or expected value of $\bar{\nu}(E)$.  By preserving the expected number of fission neutrons, the expected values of flux, reactions rates, k$_{eff}$, etc. are unbiased, but if the objective is to analyze the event-by-event nature of these reactions (e.g. simulation of neutron multiplicity counters) the microscopic behavior of the fission reaction is needed.

Through the \texttt{FMULT} input card (more details and descriptions of this input card can be found in the \MCNP\ code user manual~\cite{TechReport_2017_LANL_LA-UR-17-29981_WernerArmstrongEtAl}) the user can select from a variety of fission multiplicity treatments for both spontaneous and neutron-inducing fission reactions.  To access the \CGMF\ correlated fission model, the
\begin{Verbatim}
FMULT  METHOD=7
\end{Verbatim}
input card and option is needed.  During the initialization of \CGMF\ within the host \MCNP\ code, the \texttt{setdatapath} method is called to point \CGMF\ to the \MCNP\ location of the data files, \texttt{\$DATAPATH/cgmf}, where \texttt{\$DATAPATH} is nominally the \MCNP\ path to all data generally set as an environment variable.  Also, the \texttt{setrngcgm} method is called to point \CGMF\ to the \MCNP\ random number generator function, which will then be used everywhere a random number is needed in the initial fission fragment sampling and the Monte Carlo Hauser-Feshbach statistical de-excitation simulation of the fission fragments.

The version of \CGMF\ included in the \MCNP\ version 6.2 release is limited to the simulation of the spontaneous fission of $^{240,242}$Pu and $^{252}$Cf, and the neutron-induced fission of $^{235,238}$U and $^{239}$Pu.  Therefore, if an isotope is selected in the \MCNP\ simulation for which \CGMF\ cannot handle, the default LLNL Fission Library~\cite{TechReport_2016_LLNL_UCRL-AR-228518-REV-1_VerbekeHagmannWright} is used to produce the secondary neutron and \grays.  Again, more details on these options within the \MCNP\ code can be found in the user's manual~\cite{TechReport_2017_LANL_LA-UR-17-29981_WernerArmstrongEtAl}.

Establishing that the \CGMF\ code was integrated in the \MCNP\ code properly, a series of verification efforts using both the standalone \CGMF\ code and the integrated code were completed~\cite{TechReport_2016_LANL_LA-UR-16-23341_Rising,TechReport_2017_LANL_LA-UR-16-27710_Rising}. 

%-- FIGURE: CGMF photon vs neutron multiplicity in MCNP and standalone
\begin{figure}[ht]
\centerline{\includegraphics[width=0.75\columnwidth]{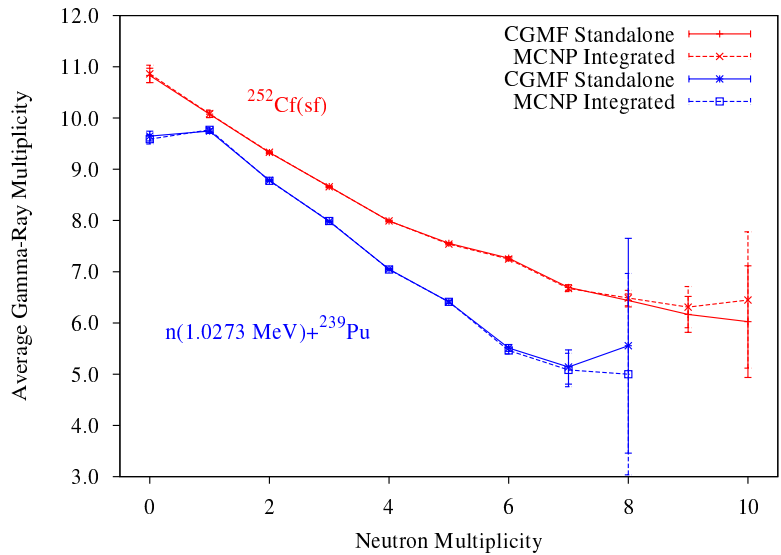}}
\caption{\label{fig:CGMFMCNPnpmult} Comparison of the standalone and integrated \CGMF\ simulated average $\gamma$-ray multiplicity as a function of neutron multiplicity for the spontaneous fission of $^{252}$Cf (red) and the $1.0273$ MeV neutron-induced fission of $^{239}$Pu (blue).}
\end{figure}
In Fig.~\ref{fig:CGMFMCNPnpmult} the event-by-event average $\gamma$-ray multiplicity as a function of neutron multiplicity is showing the integration of \CGMF\ into the \MCNP\ code provides statistically equivalent results that capture the multiplicity correlations simulated by the \CGMF\ code.  In default \MCNP\ calculations, or using other \texttt{FMULT} multiplicity treatment options, the resulting relationship in Fig.~\ref{fig:CGMFMCNPnpmult} gives a flat line, indicating the number of neutrons and $\gamma$ rays emitted in fission are uncorrelated.

%-- FIGURE: neutron-neutron angular correlations in MCNP and standalone
\begin{figure}[ht]
\centerline{\includegraphics[width=0.75\columnwidth]{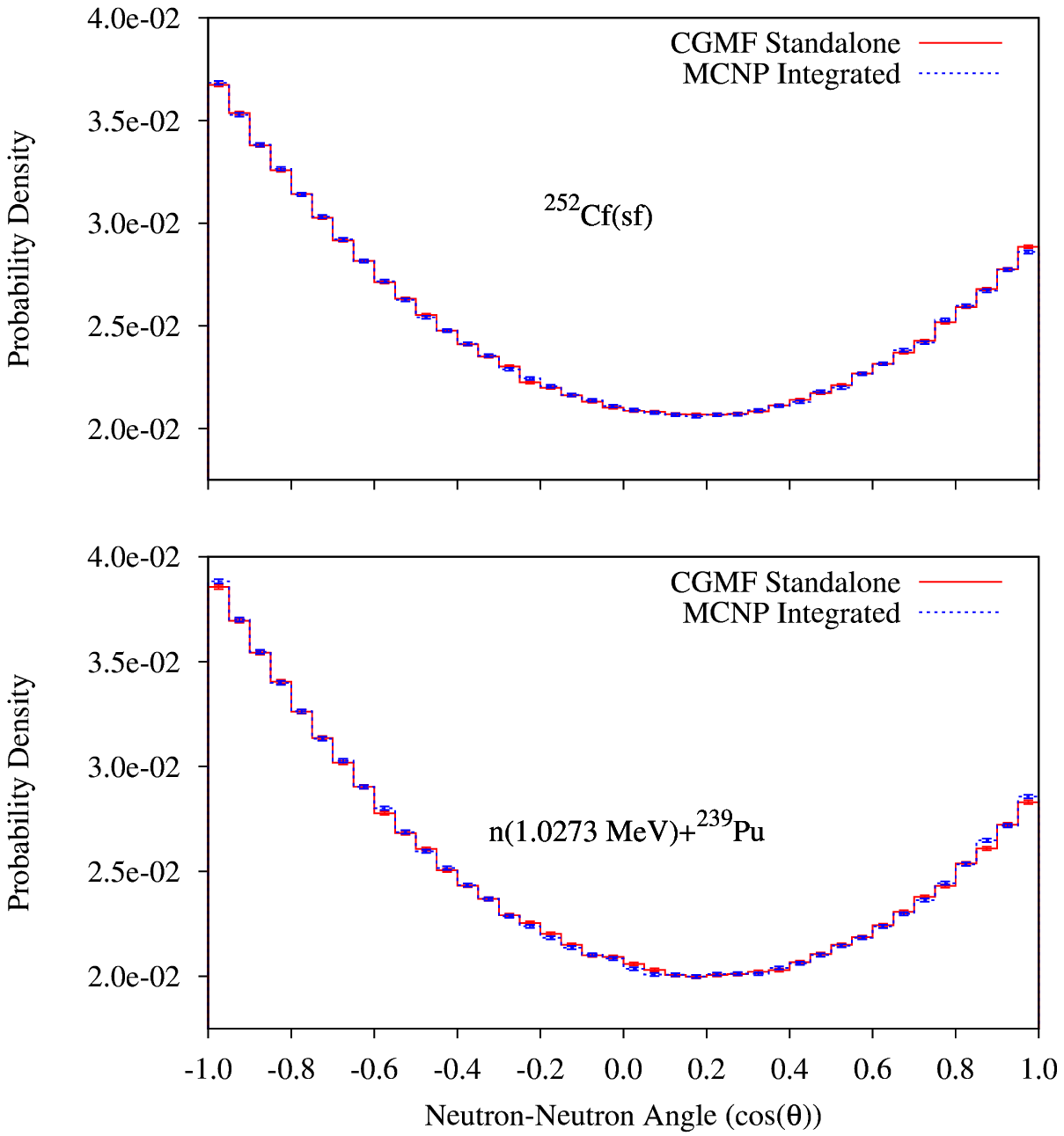}}
\caption{\label{fig:CGMFMCNPnnang} Comparison of the standalone and integrated \CGMF\ simulated neutron-neutron angular correlations for the spontaneous fission of $^{252}$Cf (top panel) and the $1.0273$ MeV neutron-induced fission of $^{239}$Pu (bottom panel).}
\end{figure} 
In Fig.~\ref{fig:CGMFMCNPnnang} the event-by-event probability density of neutron-neutron angular correlations also indicate the integration of \CGMF\ into the \MCNP\ code provides statistically equivalent results.  Again, in default \MCNP\ calculations, or using other \texttt{FMULT} multiplicity treatment options, the direction of each individual neutron emitted in fission is independently sampled from an isotropic angular distribution.
		%-- MCNP-6.2 integration
% !TEX root = ../CGMF-CPC.tex
%-- How to run CGMF
\section{Building, Testing, Installing and Running \CGMF} \label{sec:running}

With the release of \CGMF\ the build system requires the use of the \texttt{CMake} (\url{https://cmake.org/}) build process with 3.16.2 being the minimum required version of \texttt{CMake} to configure and build the code.
\texttt{CMake} is a cross-platform build system generation tool and is freely distributed at \url{https://cmake.org/download/}.  
A quick-start guide for using the \texttt{CMake} build process is given here assuming the user has the necessary C/C++ compilers installed and in the system path.  Assuming that the \CGMF\ source tree is located in the \texttt{\$CGMFPATH} directory, the Linux commands to build and test the \CGMF\ code library and utilities are:
\begin{Verbatim}
>  cd $CGMFPATH
>  mkdir build
>  cd build
>  cmake ..
>  make
>  make test
\end{Verbatim}

\noindent
This creates the static library \texttt{libcgmf.a} in the \texttt{\$CGMFPATH/build/libcgmf} directory and also creates the executable \texttt{cgmf.x} in the \texttt{\$CGMFPATH/build/utils/cgmf} directory.  The final \texttt{make test} step will execute the $30+$ \texttt{cgmf.x} fission events and yields regression tests within the \texttt{\$CGMFPATH/build/utils/cgmf/tests} directory.  Other options for building the code include:
\begin{itemize}
\item \texttt{\small  CMAKE\_BUILD\_TYPE=(Debug, RelWithDebInfo, Release [default])}
\item \texttt{\small  CMAKE\_INSTALL\_PREFIX=(/usr/local/ [default])}
\item \texttt{\small  cgmf.shared\_library=(ON, OFF [default])}
\item \texttt{\small  cgmf.tests=(ON [default], OFF)}
\item \texttt{\small  cgmf.x.MPI=(ON, OFF [default])}
\end{itemize}

\noindent
For example, to build the \CGMF\ library and executable in debug mode, with \texttt{MPI} enabled and a user-supplied install path, the \texttt{CMake} command would be:
\begin{Verbatim}
>  cmake -DCMAKE_BUILD_TYPE=Debug -Dcgmf.x.MPI=ON \
>        -DCMAKE_INSTALL_PREFIX=/path/to/install/code/and/data/
>  make
>  make test
>  make install
\end{Verbatim}

\noindent
This creates the static library \texttt{libcgmf.a} in the build directory and also creates the MPI-enabled executable \texttt{cgmf.mpi.x} in the utilities directory as described in the default build of \CGMF\ .  Additionally, by specifying the \texttt{install} target, the \texttt{/path/to/install/code/and/data/} directory will be populated with the following:
\begin{Verbatim}
   bin/cgmf.mpi.x       $ built executable
   lib/libcgmf.a        $ built static library
   cgmf/data/           $ directory of copied CGMF data files
   cgmf/include/        $ directory of copied CGMF header files
\end{Verbatim}

\noindent
Of note, this example \texttt{MPI} build of \CGMF\ requires an installation of an appropriate \texttt{MPI} library on the system, e.g. \texttt{Open MPI} (\url{https://www.open-mpi.org/}).

The \CGMF\ code either built in the \texttt{\$CGMFPATH/build/utils/cgmf} directory or installed in the \texttt{CMAKE\_INSTALL\_PREFIX/bin} directory is run on the command line and requires a few mandatory arguments and allows for a few optional arguments:
\begin{verbatim}
 -i $ZAIDt     	   [required]	1000*Z+A of target nucleus, or fissioning
                              nucleus if spontaneous fission
                            
 -e $Einc      	   [required]	incident neutron energy in MeV 
                              (0.0 for spontaneous fission)

 -n $nevents       [required] number of Monte Carlo fission events to 
                              run or to be read.  If $nevents is negative, 
                              produces initial fission fragment yields 
                              Y(A,Z,KE,U,J,p)

 -t $timeCoinc     [optional] time coincidence window for long-lived
                              isomer gamma-ray emission cutoff (in sec); 
                              if `-t -1' is given, all gamma-ray emission
                              times are given in output
                            
 -d $datapath  	   [optional]	overrides the environment variable 
                              CGMFDATA and default datapath

 -f $filename  	   [optional]	fission histories results file
                              ("histories.cgmf.0" is default, no MPI)
                            
 -s $startingEvent	[optional]	skip ahead to particular Monte Carlo event
                              (1 is default)

\end{verbatim}

\bigskip
\noindent
There are several options when it comes to providing the \texttt{cgmf.x} utility with the location of the necessary data files.  In order of priority taken by the \texttt{cgmf.x} utility, the options for specifying the path to the necessary data files are:
\begin{enumerate}
\item Use the path provided from the command line with the \texttt{-d} option
\item Use the \texttt{\$CGMFDATA} environment variable, if set
\item Use the install data path, \texttt{CMAKE\_INSTALL\_PREFIX/cgmf/data}, if installed
\item Use the source tree data path, \texttt{\$CGMFPATH/data}
\end{enumerate}

\bigskip
\noindent
As an example of running the compiled \texttt{cgmf.x} utility, the following executed from the command line
\begin{verbatim}
>  cgmf.x  -i 92235  -e 2.53e-8  -n 1000000
\end{verbatim}
simulates 1,000,000 thermal neutron-induced fission events of a $^{235}$U target.  Note that if \texttt{nevents} is negative, \CGMF\ will produce initial fission fragment yields $Y(A,Z,KE,U,J,\pi)$ but will not compute the de-excitation of those fragments. Also note that for thermal neutron-induced fission reactions, \texttt{Einc} should be set to 2.53e-8 and not 0.0; zero energy indicates the \texttt{ZAIDt} is the spontaneously fissioning nucleus.

\bigskip
\noindent
Here is an example of such an output file with its first few lines only:

\noindent
\begin{Verbatim}[gobble=2,numbers=left,numbersep=10pt,fontsize=\small,xleftmargin=5mm]
  # 98252 0.000 1e-08
  113 44 22.150 5.5 1 104.281 96.843 3 3 0
  -1700.995 2039.021 -3858.368 -1685.352 1936.768 -3712.377
  -0.448 0.100 -0.888 1.124 0.827 0.327 -0.458 1.764 0.713 -0.508 0.484 0.367 
  0.564 0.702 -0.435 1.138 -0.029 0.460 -0.887 4.822 -0.621 0.416 -0.664 2.127
  -0.460 0.135 0.878 1.466 -0.789 -0.375 0.486 0.109 -0.612 -0.487 0.623 1.372
  139 54 12.357 6.5 -1 84.767 81.862 2 2 0
  1700.995 -2039.021 3858.368 1716.558 -2016.727 3764.622
  -0.245 -0.963 -0.114 0.725 0.669 -0.355 -0.653 0.607
  -0.296 0.205 0.933 1.013 -0.041 -0.507 0.861 2.039
  -0.604 0.327 -0.726 0.410 -0.628 0.236 0.741 1.249
\end{Verbatim}

\bigskip
\noindent
Note that the line numbers are given here for sake of clarity but would not appear in the file itself.
\begin{itemize}
\item
{\bf Line 1} provides details about the fission reaction, in this case: spontaneous fission of $^{252}$Cf, \texttt{Einc} energy of $0.0$ MeV, and the time window of $1e-08$ seconds.
\item
{\bf Line 2} corresponds to the light fragment (A=113, Z=44) with an initial spin of 5.5$\hbar$ and positive parity. The pre-(104.281 MeV) and post-(96.843 MeV) neutron emission kinetic energies of the light fragment are then given. Finally, the numbers of emitted prompt fission neutrons (3), prompt fission \grays\ (3), and pre-fission neutrons (0) are given.
\item
{\bf Line 3} provides the characteristics of the fission fragment momentum components $(p_x,p_y,p_z)$ in the laboratory frame, first prior to neutron emissions, (-1700.995, 2039.021, -3858.368), then after neutron emissions, (-1685.352, 1936.768, -3712.377).
\item
{\bf Line 4} gives the directional cosines $(\hat{u},\hat{v},\hat{w})$ (-0.448, 0.100, -0.888), and energy, 1.124 MeV, of the first emitted neutron in the center-of-mass frame of the emitting fission fragment. Since in this particular example three neutrons are emitted, a total of 3$\times$(3+1)=12 numbers are provided.
\item
{\bf Line 5} gives the same information but in the laboratory frame.
\item
{\bf Line 6} provides similar information, i.e., directional cosines and energies, of all prompt \grays\ emitted. In the case of photons, only the results in the laboratory frame are provided.
\item
{\bf Lines (7-11)} repeat the same type of data but for the heavy partner fragment this time, analogous to lines (2-6) for the light fragment. In the example above, the sums of the light and heavy fragment masses (113+139) and charges (44+54) are equal to the compound fissioning system (252,98). For fission reactions at higher incident neutron energies, the possibility of multi-chance fission means that the sum of the fragment masses may be smaller than the mass of the initial fissioning nucleus. If that is the case, a new line appears below the heavy fragment data containing the characteristics-- directional cosines and energies, of those pre-fission neutrons. 
\item
Finally, this sequence of lines from (2-11) is repeated \texttt{nevents} times. 
\end{itemize}

\bigskip
\noindent
\CGMF\ provides a short summary of the average results at the end of the run. For a more in-depth statistical analysis of the results, the \CGMFtk\ python package is provided (see Section~\ref{sec:CGMFtk} for more details).

		%-- Running CGMF
% !TEX root = ../CGMF-CPC.tex
%== Python Analysis Package for CGMF Output ================================
\section{Python Analysis Package: the \texttt{CGMFtk} Toolkit}
\label{sec:CGMFtk}

By default, \CGMF\ output consists in a series of records of fission events characterized by a specific pair of fragments in initial configurations of excitation energy, spin, parity, and data characterizing the prompt neutrons and \grays\ emitted by those fragments. Although it is sometimes convenient to compute statistical means, distributions and correlations during the \CGMF\ run itself, it is often more efficient to calculate those quantities by analyzing the final output file instead. This post-run analysis can be tremendously facilitated by python tools (\texttt{matplotlib}, \texttt{numpy}, ...) freely available. We have developed a python analysis package, \CGMFtk, that provides routines to read a \CGMF\ history file and analyze the output data in powerful ways.

\CGMFtk\ is included in the \CGMF\ package.  We recommend using the \texttt{pip} python package installer \cite{pip} to install \CGMFtk\ locally.  From the command line, within the \texttt{/cgmf/CGMFtk/} folder, it can be installed via:

\bigskip
\noindent
\texttt{pip3 install .} 

\bigskip

\noindent Now, the classes within \CGMFtk\ can be imported like any other python package,

\bigskip
\noindent \texttt{from CGMFtk import histories as fh} \\
\noindent \texttt{from CGMFtk import yields as yld} 
\bigskip

There are two classes within \CGMFtk.  The first is $\mathtt{CGMFtk.histories}$ which creates a fission history object through $\mathtt{CGMFtk.histories.Histories(filename)}$. The required argument is the name of the \CGMF\ history file. In addition, the optional keyword \texttt{nevents} can be included to limit the number of events read in from the history file.  The second class is $\mathtt{CGMFtk.yields}$ which creates a yield history object from the result of \CGMF\ runs when only the yields are calculated; this corresponds to the  use of a negative number of events in the  $\texttt{-n}$ command line option.  The object is created through $\mathtt{CGMFtk.yields.Yield(filename)}$.  Again, the name of the yields file is required, and the number of events to read is an optional keyword.  

An example python notebook is provided as part of the \CGMF\ package, giving some examples of the methods implemented in these classes.  In addition, the examples outlined in Section~\ref{sec:applications} below make use of some of the features of the \CGMFtk\ toolkit.
	%-- Python analysis package
% !TEX root = ../CGMF-CPC.tex
%-- Applications
\section{Applications}
\label{sec:applications}

The \CGMF\ code is a very versatile tool for predicting and analyzing nuclear fission experiments. We provide below a few examples of how \CGMF\ can be used to predict distributions and correlations among emitted prompt neutrons and \grays, compute isomeric ratios in fission fragments, infer fission product yields using \gray\ spectroscopy, and compute the time evolution of the prompt \grays\ emitted from isomers populated in the fission fragments. This represents a very small subset of all the interesting uses and applications that can be made using \CGMF\ and its analysis toolkit \CGMFtk.

%-------------------------------------------------------------------------------------------------------------
%-- Applications: Prompt Fission Neutron and Gamma Multiplicity Distributions
%-------------------------------------------------------------------------------------------------------------
\subsection{Prompt Fission Neutron and \g\ Multiplicity Distributions}

One of the easiest data that can be pulled out of \CGMF\ history files are the multiplicity distributions of both neutrons, $P(\nu)$, and \grays, $P(N_\gamma)$. Accurately calculating the average number of prompt fission neutrons, \nubar, requires about 50,000 fission events only, while computing accurately its distribution $P(\nu)$ requires many more events, depending on the precision one requires on the tail of the distribution. The results converge even faster for \grays\ which are emitted in greater number than neutrons.

When comparing with experimental data, one should be aware of some complications due to the detector energy threshold and, for the \grays, the fission time coincidence window used in the particular experiment as the average \gray\ multiplicity is very sensitive to this number. 

\begin{figure}[ht]
\centerline{
\includegraphics[width=0.5\columnwidth]{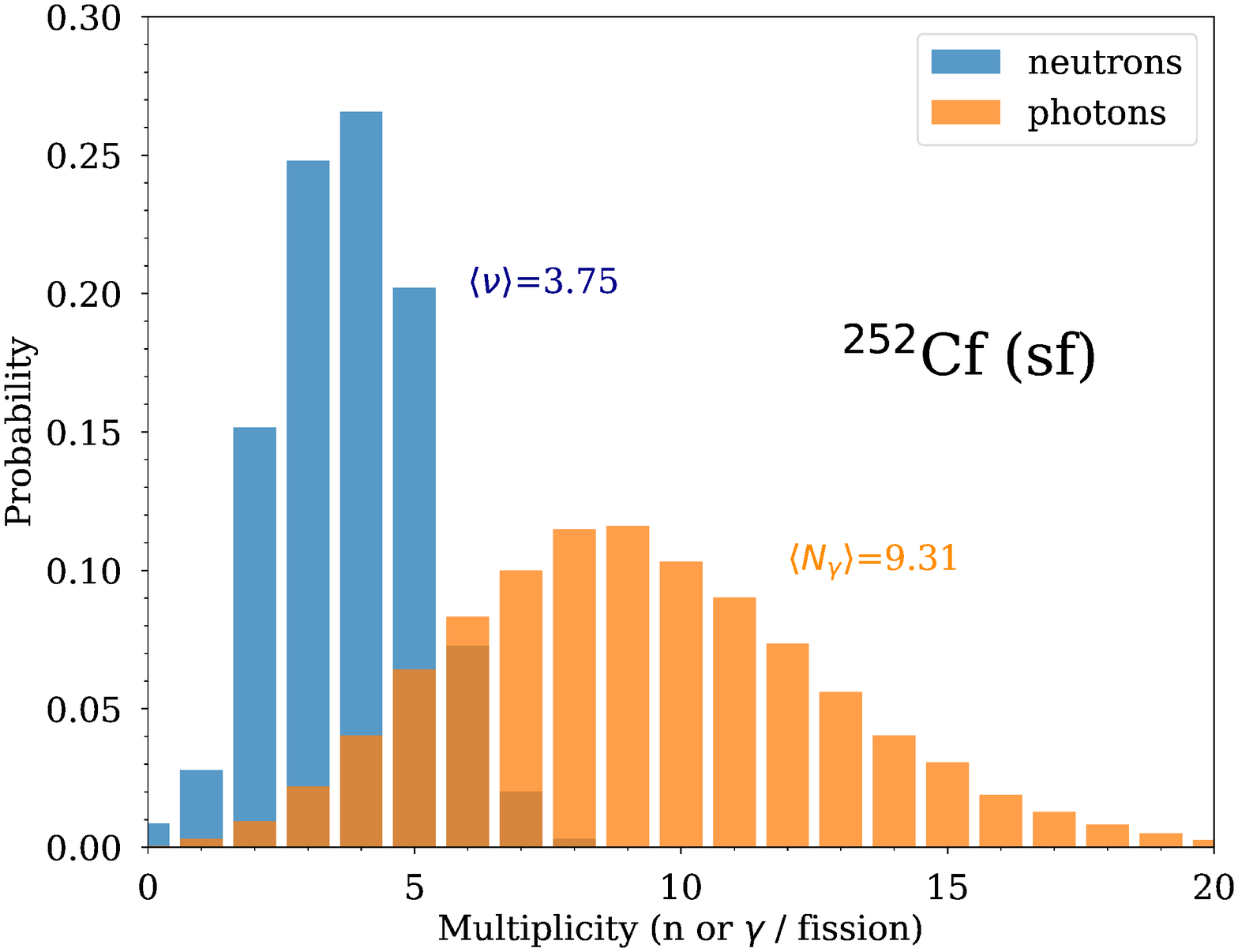}
\includegraphics[width=0.5\columnwidth]{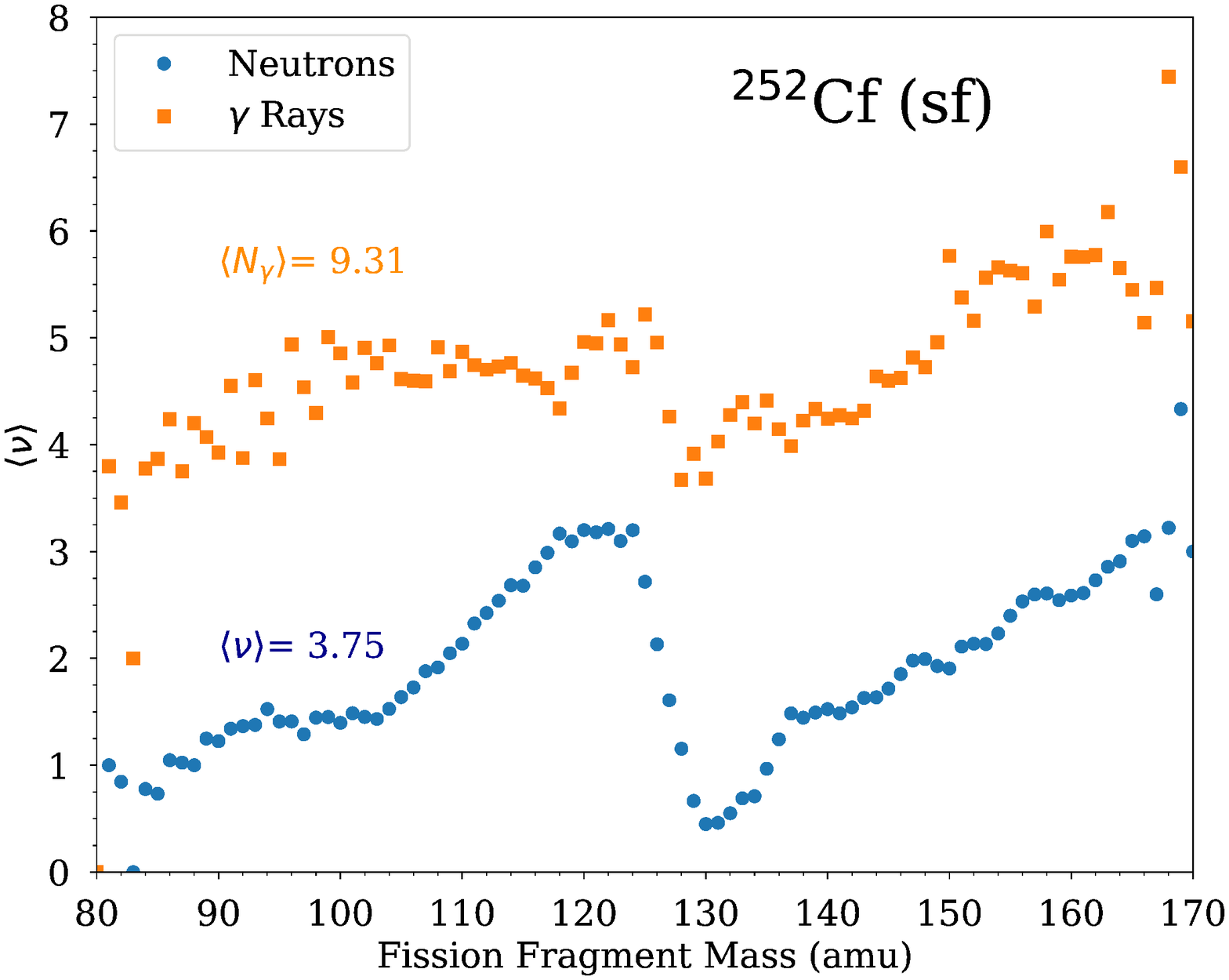}
}
\caption{\label{fig:nu-nug}(a) Prompt neutron and \gray\ multiplicity distributions obtained by using the built-in functions \texttt{h.Pnu()} and \texttt{h.Pnug()} after reading the \CGMF\ histories with \texttt{h=fh.Histories().} (b) The average neutron and \gray\ multiplicities as a function of the fission fragment mass are calculated using \CGMFtk\ and the built-in functions \texttt{h.nubarA()} and \texttt{h.nubargA()}.}
\end{figure}

After executing \CGMF\ for a particular fission reaction and number of events, let's say:
\begin{eqnarray}
\texttt{./cgmf.x -i 98252 -e 0.0 -n 1000000}, \nonumber
\end{eqnarray}
which corresponds to the spontaneous fission of $^{252}$Cf with one million fission events, the output history file can be read as follows
\begin{eqnarray}
\texttt{h = fh.Histories ("histories.cgmf")}
\end{eqnarray}
The neutron and \gray\ multiplicity distributions can then be easily extracted using
\begin{eqnarray}
\texttt{h.Pnu()} \nonumber \\
\texttt{h.Pnug()}, \nonumber
\end{eqnarray}
which produce two \texttt{numpy} arrays with pairs of values ($\nu$,$P(\nu)$). Those arrays can then be easily plotted using \texttt{matplotlib} (used here) or any other plotting software. The result is shown in Fig.~\ref{fig:nu-nug} (a). As mentioned earlier, $\langle N_\gamma\rangle$ depends significantly on the time coincidence window used to account for prompt fission \g\ rays.

%-- Figure: P(nu) and P(nu_gamma)

Note that the average neutron multiplicity \nubar\ per {\it fission event} is obtained using the function \texttt{h.nubartot()}, not to be confused with the function \texttt{h.nubar()}, which provides the average neutron multiplicity per {\it fission fragment}.

%-------------------------------------------------------------------------------------------------------------
%-- Applications: Prompt Particle Correlations
%-------------------------------------------------------------------------------------------------------------
\subsection{Prompt Particle Correlations}

What makes the \CGMF\ code particularly interesting is that it provides a complete reconstruction of the fission events, characterized by the fission fragment masses, charges, and kinetic energies, their initial conditions, i.e., right after scission, in energy, spin, and parity, and all the emitted prompt neutron and \gray\ energies, multiplicities, and angular distributions. Some correlations can easily be drawn using built-in functions from \CGMFtk. For instance, calculating the average neutron and \gray\ multiplicities as a function of the fragment mass simply requires:
\begin{eqnarray}
\texttt{h.nubarA()} & \mbox{for neutrons} \nonumber \\
\texttt{h.nubargA()} & \mbox{for \grays} \nonumber
\end{eqnarray}
The result is shown in Fig.~\ref{fig:nu-nug} (b). Again, the calculated $\langle N_\gamma\rangle(A)$ depends somewhat on the time coincidence window imposed around the fission event. 

\begin{figure}[ht]
\centerline{\includegraphics[width=0.75\columnwidth]{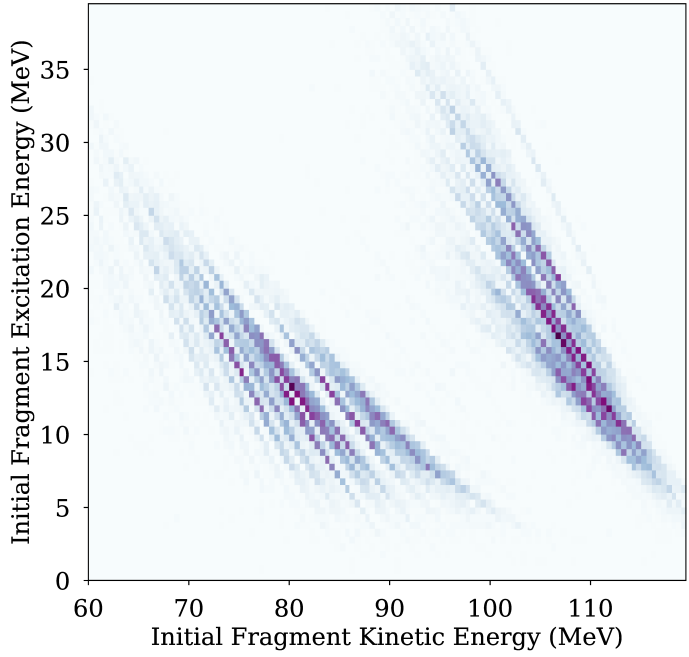}}
\caption{\label{fig:U-KE}The relation between the initial fragment excitation energy and the fragment kinetic energy can be plotted using the Matplotlib function \texttt{hist2d()}, after retrieving those quantities using \texttt{h.getKEpre()} and \texttt{h.getU()}.}
\end{figure}

Two-dimensional correlation plots can also be easily made by invoking the \texttt{matplotlib} Python package and its two-dimensional histogram function. For instance, let us plot the initial fission fragment excitation energy as a function of the fragment kinetic energy before neutron emission:
\begin{eqnarray}
\texttt{plt.hist2d(h.getKEpre(),h.getU())} \nonumber
\end{eqnarray}
with the resulting plot shown in Fig.~\ref{fig:U-KE}.

%-- Figure: (U,KE) 2d histogram

%-------------------------------------------------------------------------------------------------------------
%-- Applications: Fission Yields and Gamma-Ray Spectroscopy
%-------------------------------------------------------------------------------------------------------------
\subsection{Fission Yields and \gray\ Spectroscopy}

Towards the end of the de-excitation process, a fission fragment will typically decay by emitting \grays\ between excited discrete levels until it reaches its ground state or a long-lived isomeric state. These levels are provided in the \texttt{cgmfDiscreteLevels-2015.dat} file, described in Sec.~\ref{sec:structure}, and derived from the RIPL library. Discrete transitions are unique for each nucleus, meaning that one can relate the intensity of particular \gray\ transitions to the abundance of the particular nucleus attributed to it. Since those \grays\ are emitted are neutron emission, the abundance or {\it yield} of this fragment an correspond to a multitude of pre-neutron emission fission fragments which decay according to a certain neutron multiplicity distribution P$(\nu)$, leading to the same post-neutron emission fragment. 

\begin{figure}[ht]
\centerline{\includegraphics[width=\columnwidth]{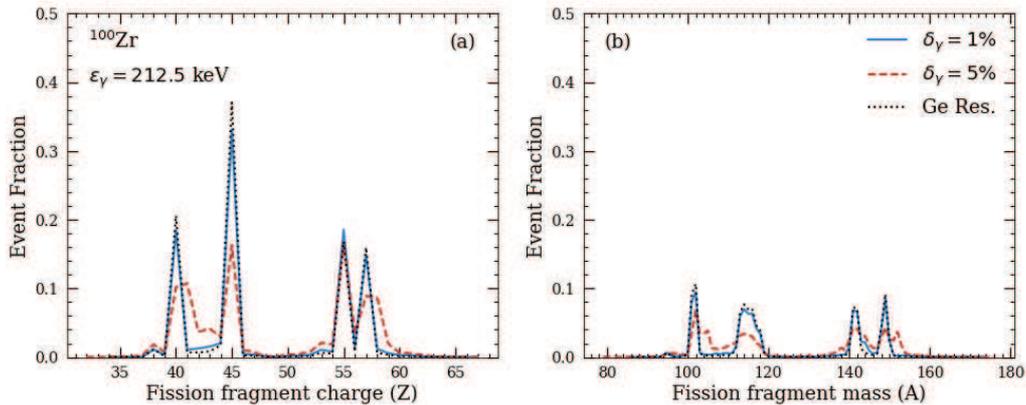}}
%\centerline{\includegraphics[width=0.8\columnwidth]{figs/blank.png}}
\caption{\label{fig:Gspec}Histograms of the contributing fission fragments in both charge (a) and mass (b) producing the $^{100}$Zr $2^+\rightarrow 0^+$ transition with an energy $\epsilon_\gamma = 212.50$ keV. Three different energy resolutions, corresponding to $1\%$ (solid), $5\%$ (dashed), and a typical Ge resolution~\cite{Owens:1985} (dotted) are provided. This \texttt{CGMF} calculation utilized 1 million events of $^{252}$Cf(sf).}
\end{figure}

This technique, referred to as \gray\ spectroscopy, can be facilitated using our \CGMF\ toolkit, \texttt{CGMFtk}, which provides the following routine:
\begin{eqnarray}
\texttt{h.gammaSpec($\epsilon_\gamma$,$\delta_{\epsilon_\gamma}$,post=True)}
\end{eqnarray}
that lists all post-neutron emission fission fragments producing a \gray\ with energy $\epsilon_\gamma$, within the energy window 
[$\epsilon_\gamma - \delta_{\epsilon_\gamma},\epsilon_\gamma + \delta_{\epsilon_\gamma}$], where $\delta_{\epsilon_\gamma}$ plays the role of an energy resolution. The results of this tool can be easily plotted as in Fig.~\ref{fig:Gspec}, to determine which fission fragments are contributing to particular \grays.

%-- FIGURE: Gamma-ray spectroscopy on a 2+ --> 0+ transition in Zr100.

It is important to note that \CGMF\ does not {\it predict} the presence of discrete levels, but instead take those from the ENSDF or RIPL libraries. Likewise, branching fractions and internal conversion coefficients are taken from the same data file. What \CGMF\ does however is compute the neutron and \g\ emission probabilities prior to feeding particular discrete states. 

%-------------------------------------------------------------------------------------------------------------
%-- Applications: "Late" Prompt Gamma Emissions
%-------------------------------------------------------------------------------------------------------------
\subsection{``Late" Prompt \g\ Emissions}

\begin{figure}[ht]
\centerline{\includegraphics[width=0.65\columnwidth]{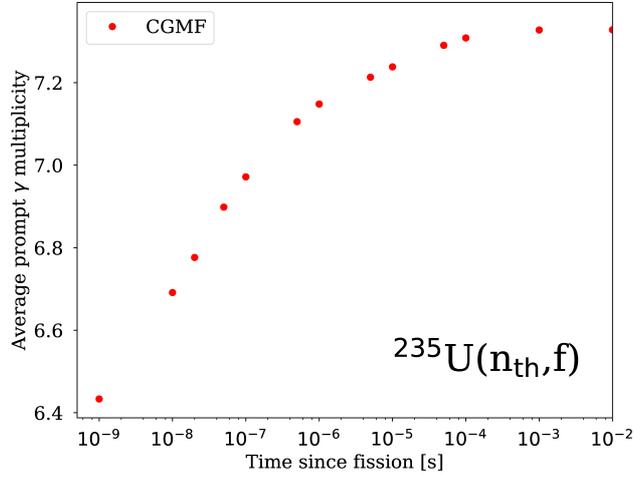}}
\caption{\label{fig:ng-vs-t}The average number of \grays\ as a function of time since fission for the $^{235}$U($n_\mathrm{th},f$) reaction. The threshold for the detection of the \grays\ has been set to 100 keV for this calculation.}
\end{figure}

The prompt \grays\ are emitted from the fission fragments, in coincidence with a fission event, and before any $\beta$ decay. Experimentally, a {\it time coincidence window} is typically defined from a few nanoseconds to tens or even hundreds of nanoseconds of a fission event. Although most of the prompt \grays\ are emitted after neutron emission, the existence of isomeric states in the fission fragments produces ``late'' (or ``delayed'') prompt \g\ emissions, and thus a dependence on the time-coincidence window of the \gray\ observables. In some cases, like the average prompt fission \g\ spectrum, the dependence is fairly weak, but for the average prompt fission \gray\ multiplicity the effect can be as large as 5 to 13\%~\cite{Talou:2016}, as isomeric states with a variety of lifetimes are populated. The time dependence of the average prompt \g\ multiplicity is presented in Fig. \ref{fig:ng-vs-t} for the fission of $^{235}$U induced with thermal neutrons.  This dependence can be calculated with the $\texttt{nubarg}$ routine by including the option $\texttt{timeWindow}$:
\begin{eqnarray}
\texttt{h.nubarg(timeWindow=True)}.
\end{eqnarray}
The option, $\texttt{timeWindow}$, takes either a Boolean or a list of times (in seconds) at which to calculate the \gray\ multiplicity.  In addition, the keyword $\texttt{Eth}$ can be included to set a lower bound on the \gray\ energy.

%-- FIGURE: Time dependence of the average prompt fission gamma multiplicity

The nuclear structure data file used in \CGMF, and extracted from the \RIPL\ database~\cite{RIPL3}, contains information about the lifetime of some discrete states, and this information is used to sample a time after which the state decays, assuming an exponential decay probability as a function of time given by
\begin{equation}
P(t)\propto 1-\exp \left(-t \log(2)/\tau_{1/2}\right),
\end{equation}
where $\tau_{1/2}$ is the half-life of the particular state. If no half-life is available, we assume that the electromagnetic transition is instantaneous. For states with non-zero half-lives, if the sampled time is larger than the time coincidence window, the \g\ cascade is stopped, as an experiment with that particular coincidence window would not be able to observe further \gray\ emissions. The reliability of the simulation relies on the reliability of the nuclear structure data, and, thus, it is likely that future updates of the \texttt{RIPL} database will change this time dependence.

%\textcolor{red}{In this section, it would be good to have an example of how this is extracted from \CGMF\ runs. See other similar subsections above.}

%-- FIGURE: Number of gamma rays emitted as a function of time since fission
\begin{figure}[ht]
\centerline{
\includegraphics[width=0.5\columnwidth]{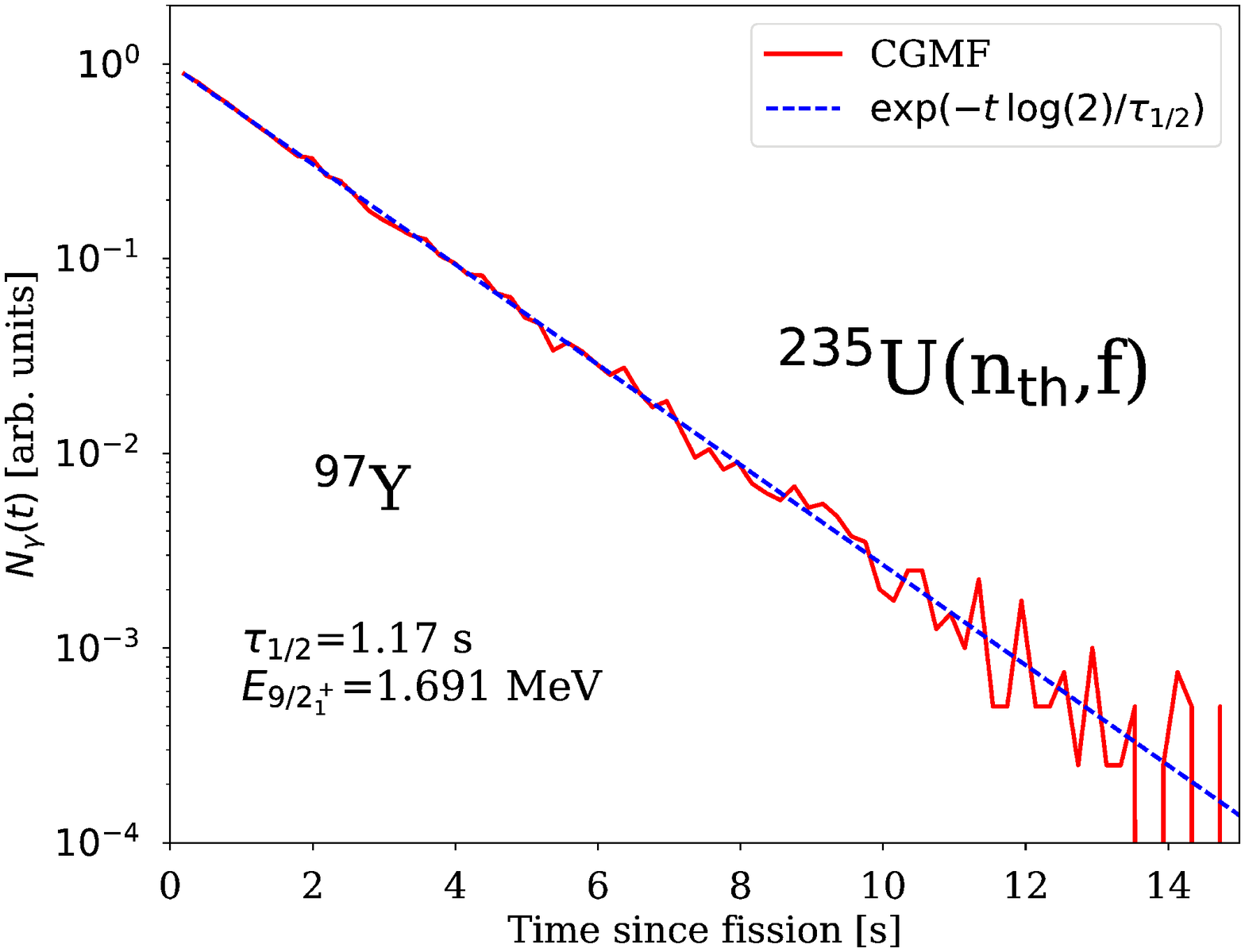}
\includegraphics[width=0.5\columnwidth]{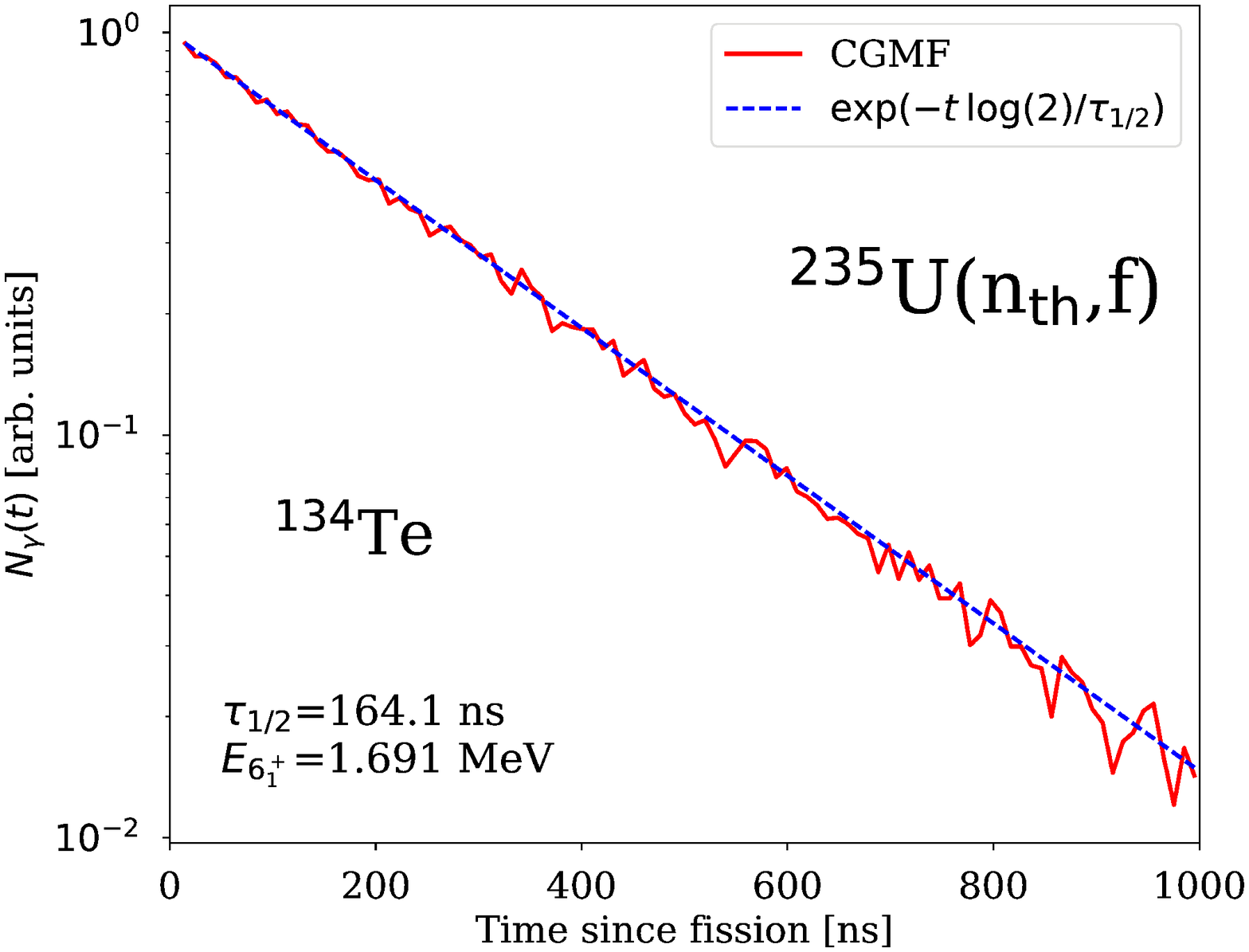}
}
\caption{\label{fig:decay-states}The number of \grays\ emitted as a function of time from $^{97}$Y (left panel) and $^{134}$Te (right panel) fission fragments produced in the fission of $^{235}$U induced with thermal neutrons.}
\end{figure}

In Fig. \ref{fig:decay-states}, we illustrate the decay of isomeric states in two fission fragments produced in the $^{235}$U($n_\mathrm{th},f$) reaction after neutron emission, by plotting (in arbitrary units) the number of \grays\ emitted from the fragments as a function of time. The time dependence of the number of \grays\ emitted is dominated by the decay of one particular isomeric state in each fragment, although other shorter lived states may exist. \CGMF\ properly reproduces the exponential decay of the states for the two fission fragments, which span entirely different time scales.

The default time coincidence window in \CGMF\ is set at 10 ns, but the option \texttt{-t} allows the users to set a custom time coincidence window. If the user provides -1.0 for the time coincidence window, this is equivalent to an infinite time window, so all possible \gray\ emissions are recorded. 

\CGMFtk\ provides a function to calculate the number of \grays\ emitted as a function of time since fission when the history file is created with the option \texttt{-t -1.0}, using 
\begin{eqnarray}
\texttt{h.gammaMultiplicity($t_{min}$,$t_{max}$,$A_F$,$Z_F$)}
\end{eqnarray}
\noindent where $t_{min}$ and $t_{max}$ define a time window around the half-life of the fission fragment specified by $A_F$ and $Z_F$.
%We provide the python notebook \texttt{testTimeImplementation.ipynb}, which has the capability of reading from the history file created with the option \texttt{-t -1.0} only the \g-ray information, including the time the \g-ray was produced after fission. This information can be then used to obtain different quantities that depend on the age of the \grays, like the average PFG multiplicity, spectrum, decay of fission isomers, and isomeric ratios. The user can provide a custom history file or the one provided with the package, and generate the plots in Figs. \ref{fig:ng-vs-t} and \ref{fig:decay-states}.

%-------------------------------------------------------------------------------------------------------------
%-- Applications: Isomeric Ratios
%-------------------------------------------------------------------------------------------------------------
\subsection{Isomeric Ratios}

A consequence for the existence of long-lived or isomeric states in fission fragments is the access to  isomeric ratios, which are directly measurable for some isotopes. As in our previous paper~\cite{Stetcu:2013}, we define the isomer ratio as the ratio of the yield of the highest-spin state (either the isomer itself or the ground state) to the total production of the isotope:
\begin{eqnarray}
{\rm IR}=\frac{Y^{\rm isomer}_{\rm high-spin}}{Y_{\rm tot}}.
\end{eqnarray}
When the spins of the isomer and the ground-state differ significantly, the prediction of the isomeric ratio is a test of the initial spin distribution in the fragments leading to their production. A high-spin state can only be strongly populated if the initial spin distribution is high enough for instance. However, as also noted in~\cite{Stetcu:2013}, the extraction of the spin is strongly model-dependent, as it is sensitive to a host of quantities like nuclear level density, neutron optical model, \gray\ strength function, discrete levels and branching ratios for electromagnetic transitions between discrete levels. 

Running \CGMF\  using the option \texttt{-t -1} produces a history file that lists all the \gray\ ages, i.e., when they are emitted, with no constraints on \gray\ emission based on the time since fission. This history file can then be used to calculate isomeric ratios using:
\begin{eqnarray}
\mathtt{h.isomericRatio(T_{\rm thres},A_F,Z_F,J_m,J_{gs})},
\end{eqnarray}
\noindent where $A_F$ and $Z_F$ are the fragment mass and charge, $J_m$ is the spin of the isomeric state, and $J_{gs}$ is the spin of the ground state of the nucleus. In order to obtain a reliable result for the isomeric ratio, one needs to provide as input for the threshold time, $T_{\rm thres}$, a value much smaller than the lifetime of the isomer, and much larger than any other lifetime for other states. The isomeric ratio will not depend on the choice of the threshold time if the time is chosen much smaller than the lifetime. It is recommended that the user performs a series of tests to ensure that the isomeric ratio is independent of the threshold. In Table \ref{table:ir}, we provide a few examples of isomeric ratios obtained with \CGMF\, and the threshold time used in the calculation. Finally, the user should be aware that a large number of fission events may have to be generated, depending on the abundance of the particular fission fragment whose isomer is analyzed (in our history file we provide 1.4M fission events).

\begin{table}[h]
\caption{Isomeric ratios computed with \texttt{CGMFtk} for the $^{235}$U(n$_\mathrm{th}$,f) reaction. We provide the lifetime for each fission fragment, the spin of the ground and isomeric states, and the threshold time used in the calculation. \label{table:ir}}
\begin{tabular}{ccccccc}
\hline \hline
Nucleus &  Spin Isomer & Spin Ground State & Lifetime (s) & $T_\mathrm{threshold} $(s) & Isomeric Ratio\\
\hline
$^{83}$Se & $1/2^-$ & $9/2^+$ & 70.1 & $0.1$ & 0.79 \\
$^{99}$Nb & $1/2^-$ & $9/2^+$ & 150 & 1 & 0.91 \\
$^{133}$Te & $11/2^-$ & $3/2^+$& 917.4 & 1 & 0.33 \\
$^{135}$Xe & $11/2^-$ & $3/2^+$& 3324 & 1 & 0.55 \\
\hline\hline
\end{tabular}
\end{table}

In comparison with Fig. 4 in~\cite{Stetcu:2013}, we see significant deviations for $^{133}$Te, where the current result is about half the value reported earlier. This is very likely due to the current treatment of the fission yields and/or discrete data, which have changed considerably from the time Ref.~\cite{Stetcu:2013} was published. 
	%-- Applications
% !TEX root = ../CGMF-CPC.tex
%-- CONCLUSION ---------------------------------------------------------------
\section{Conclusion}
\label{sec:conclusion}

This paper describes the \CGMF\ code that follows the de-excitation of fission fragments through the emission of prompt fission neutrons and \grays\ on an event-by-event basis. By following those successive emissions every step of the decay, the \CGMF\ main output is a history file that contains the details in energy, angle and time of emission for each of the emitted particles. A wide range of distributions and correlations pertaining to those neutrons and \grays\ can then be inferred through straightforward statistical analyses. Intricate $n$-$n$, $n$-\g\ and \g-\g\ correlations can be extracted to study complex signatures of the nuclear fission process. The event-by-event nature of the output also allows its use in detector response simulations.

The applicability of \CGMF\ is currently limited by the availability and accuracy of the pre-neutron fission fragment distributions in mass, charge, and kinetic energy. Spontaneous fission ($^{238,240,242,244}$Pu and $^{252,254}$Cf) and neutron-induced ($^{233,234,235,238}$U, $^{237}$Np, and $^{239,241}$Pu) fission reactions are available up to 20 MeV incident energy. 

The physics models implemented in \CGMF\ are similar, if not identical, to the ones present in modern nuclear reaction codes, e.g., \COH, and used to develop nuclear data files in existing evaluated libraries, e.g., ENDF/B-VIII.0~\cite{Brown:2018}. Note that \CGMF\ has not been optimized to provide prompt fission neutron spectrum calculations as accurate as one would wish. In particular, neutron spectra calculated with \CGMF\ tend to be softer than what is presently included in most evaluated data files. The use of \CGMF\ results for PFNS and/or for any calculation that strongly depends on the energies of the emitted neutrons could be biased.
	%-- Conclusion

%== ACKNOWLEDGMENTS =====================================================
\section*{Acknowledgments}

This work was supported by the Office of Defense Nuclear Nonproliferation Research \& Development (DNN R\&D), National Nuclear Security Administration, US Department of Energy, and was performed at Los Alamos National Laboratory under contract No.~89233218CNA000001.

%== REFERENCES ============================================================
\nocite{*}
\bibliographystyle{cpc}
\bibliography{references-cgmf}

\end{document}